\documentclass[prx, reprint, onecolumn, superscriptaddress,
notitlepage, nofootinbib, longbibliography,
floatfix]{revtex4-2}
\usepackage{graphicx} 
\usepackage{amsmath, amssymb, amsthm}
\usepackage{mathrsfs}
\usepackage{tensor}
\usepackage[utf8]{inputenc}
\usepackage{mathtools}
\usepackage{tikz}
\usepackage{tikz-cd}
\usetikzlibrary{calc,topaths,decorations,decorations.pathmorphing,arrows,decorations.markings,cd}
\usepackage{xcolor}

\newtheorem{example}{Example}

\DeclareMathOperator{\Tr}{Tr}
\DeclareMathOperator{\End}{End}
\DeclareMathOperator{\Inhom}{\underline{\hom}}
\DeclareMathOperator{\Rep}{Rep}
\DeclareMathOperator{\cRep}{\mathsf{Rep}}

\DeclareMathOperator{\Vecc}{\mathsf{Vec}}
\DeclareMathOperator{\Zc}{\mathcal{Z}}
\DeclareMathOperator{\bimod}{Bimod}

\DeclareMathOperator{\Endo}{End}
\DeclareMathOperator{\Inn}{Inn}
\DeclareMathOperator{\id}{id}
\DeclareMathOperator{\ev}{ev}
\DeclareMathOperator{\irrep}{irrep}

\colorlet{algebra}{blue!70!green}
\colorlet{algebra2}{black!30!red}
\colorlet{module}{orange}
\colorlet{simple}{black!20!green}

\newcommand{\LeftModuleAction}[1]{
    \begin{scope}[shift={#1}]
        \fill[module] (0,0.12) to[in=60, out=180] (-0.12,0) -- (0,0) -- cycle;
    \end{scope}
}
\newcommand{\RightModuleAction}[1]{
    \begin{scope}[shift={#1}]
        \fill[module] (0,0.12) to[in=120, out=0] (0.12,0) -- (0,0) -- cycle;
    \end{scope}
}
\newcommand{\insertion}[2]{
    \begin{scope}[shift={#1}]
        \fill[simple] 
            (0,0) -- (-0.06,-0.15) -- (0,-0.15) -- (0.06,-0.15) -- cycle;
        \draw[thin] (0.06,-0.15) -- (0,0) -- (-0.06,-0.15);
        \draw[] (0, -0.06) node [right] {\footnotesize ${#2}$};
    \end{scope}
}
\newcommand{\projection}[2]{
    \begin{scope}[shift={#1}]
        \fill[simple] 
            (0,0) -- (-0.06,0.15) -- (0,0.15) -- (0.06,0.15) -- cycle;
        \draw[thin] (0.06,0.15) -- (0,0) -- (-0.06,0.15);
        \draw[] (0, 0.06) node [right] {\footnotesize ${#2}$};
    \end{scope}
}
\newcommand{\multiplication}[1]{
    \begin{scope}[shift={#1}]
        \fill[red] (0,0) circle(0.04);
        \draw[] (0,0) circle(0.04);
    \end{scope}            
}
\newcommand{\unit}[1]{
    \begin{scope}[shift={#1}]
        \fill[white] (0,0) circle(0.04);
        \draw[] (0,0) circle(0.04);
    \end{scope}
}

\begin{document}
\preprint{APS/123-QED}
\title{Classification of intrinsically mixed $1+1$D non-invertible $\Rep(G) \times G$ SPT phases}

\author{Youxuan Wang}
\affiliation{Department of Physics, The Hong Kong University of Science and Technology, Clear Water Bay, Hong Kong, China}

\begin{abstract}
We classify $1{+}1$d bosonic SPT phases with non-invertible symmetry $\Rep(G)\times G$, equivalently the
fusion-category symmetry $\mathcal H=\cRep(G)\boxtimes\Vecc_G$.
Focusing on \emph{intrinsically mixed} phases (trivial under either factor alone), we use the correspondence
between $\mathcal H$-SPTs, $\mathcal H$-modules over $\Vecc$, and fiber functors $\mathcal H\to\Vecc$
\cite{gelakiTensorCategories2015,inamura202411dsptphasesfusion} to obtain a complete classification:
such phases are parametrized by $\phi\in \End(G)/\Inn(G)$.
For each $\phi$ we identify the associated condensable (Lagrangian) algebra $\mathcal A_\phi$ in the bulk
$\Zc(\mathcal H)\simeq \mathcal D_G^2$.
We further provide an explicit lattice realization by modifying Kitaev’s quantum double model with a domain
wall $\mathcal B_\phi$ and smooth/rough boundaries \cite{KITAEV20032,liGappedBoundariesKitaevs2025}, and then
contracting to a 1D chain, yielding a (possibly twisted) group-based cluster state
\cite{PhysRevX.15.011058} whose ribbon-generated symmetry operators encode the same $\phi$.
\end{abstract}
\maketitle

\section{Introduction}
Symmetry-protected topological (SPT) phases are gapped short-range-entangled phases that are adiabatically
connected to trivial product states once the protecting symmetry is forgotten, but become distinct in the
presence of a global symmetry. In one spatial dimension, SPT order is equivalently characterized by the
absence of intrinsic bulk topological order together with robust boundary degrees of freedom, or by the
impossibility of disentangling the ground state by any symmetry-preserving finite-depth local unitary
circuit. For bosonic on-site \emph{group} symmetries, 1D SPT phases are classified by projective
representations---equivalently by the group cohomology class in \(H^{2}(G,U(1))\)---as understood from
matrix-product-state approaches and related viewpoints
\cite{Haldane1983ContinuumHeisenberg,AKLT1987,Pollmann2010SymmetryProtection1D,Schuch2011ClassifyingMPS,
Chen2011Classification1DSPT,Chen2013SPTGroupCohomology,Hastings2007QuasiAdiabatic,LevinGu2012BraidingSPT}.

More recently, the notion of symmetry itself has been broadened by the perspective that symmetries are
implemented by topological defect operators. In quantum field theory this includes higher-form symmetries
generated by higher-codimension topological operators \cite{Gaiotto2015GeneralizedGlobalSymmetries,
Kapustin2014Coupling,McGreevy_2023,cordova2022snowmasswhitepapergeneralized}. Beyond invertible defects, one
encounters \emph{non-invertible} symmetries, whose defects fuse according to a fusion algebra rather than a
group law \cite{Fendley2012Parafermions,Fendley2019FreeParafermions,ElseNayak2014ClassicalNonAbelian,
Aasen2016TopologicalDefects,thorngrenFusionCategorySymmetry2024}. In 1+1D such symmetries are naturally
encoded by a fusion category \(\mathcal C\), and gapped phases with \(\mathcal C\) symmetry admit a
categorical/topological description in terms of module categories, fiber functors, and the bulk Drinfeld
center \(\mathcal Z(\mathcal C)\) (the ``Symmetry TFT'' viewpoint)
\cite{gelakiTensorCategories2015,ENO2002,Ostrik2003ModuleCategoriesDouble,Inamura2022LatticeModelsFusionCategorySym,
inamura202411dsptphasesfusion}. This framework is particularly well-suited for constructing microscopic
realizations and identifying explicit nonlocal observables, as exemplified by recent group-based cluster
state constructions with non-invertible \(\Rep(G)\times G\) symmetry \cite{SeifnashriShao2024ClusterStateNoninvertibleSPT,
PhysRevX.15.011058,Brell2015GeneralizedClusterStatesFiniteGroups} and by lattice realizations of gapped
boundaries/domain walls in quantum double models \cite{KITAEV20032,Beigi_2011,liGappedBoundariesKitaevs2025}.

Recent progress has made categorical SPT phases with group-theoretical symmetry data increasingly concrete at
the lattice level. In particular, Ref.~\cite{PhysRevX.15.011058} provides an explicit 1D stabilizer
construction realizing a nontrivial \(G\times \Rep(G)\)-protected phase, exhibiting protected edge degrees of
freedom, string order, and a quantized charge--flux response. From a complementary and more systematic
perspective, Ref.~\cite{inamura202411dsptphasesfusion} formulates \(1{+}1\)d SPT phases with finite fusion-category
symmetry \(\mathcal C\) in the MPO--MPS framework, relating phases to fiber-functor data, deriving an interface
algebra whose representation theory enforces degenerate interface modes, and identifying \(S^{1}\)-families via
a non-Abelian Thouless-pump invariant \cite{Bultinck2017AnyonsMPS,Bultinck2017MPOSymmetry,
Williamson2016MatrixProductSymmetries,Verstraete2008MPSReview}.

These developments motivate moving beyond ``factorized'' settings, where nontriviality can already be detected
after restricting to a single symmetry factor. For the product symmetry
\(\mathcal C=\cRep(G)\boxtimes \Vecc_{G}\), there is a distinguished class of \emph{intrinsically mixed} phases:
they become trivial upon restriction to either the charge sector \(\cRep(G)\) or the flux sector \(\Vecc_{G}\),
yet remain nontrivial under the full product symmetry. Physically, such phases isolate response/obstruction
data that depends essentially on the interplay between charge and flux defects, reminiscent of the
charge--flux structure underlying quantum doubles and their boundaries/domain walls
\cite{DijkgraafWitten1990,DPR1990,DeWildPropitius1995TopologicalInteractions,BaisSlingerland2009Condensation,
Kong2014AnyonsGappedBoundaries,BarkeshliBondersonChengWang2019SymmetryDefects}.
Accordingly, we take intrinsically mixed phases as a minimal arena in which \(\cRep(G)\) and \(\Vecc_G\) must be
treated on equal footing, and where genuinely non-factorizable categorical-SPT phenomena can be cleanly
identified.

We give a complete classification of \emph{intrinsically mixed} \(\Rep(G)\times G\) SPT phases, showing that they are
parameterized by endomorphisms \(\phi\in \End(G)/\Inn(G)\) (equivalently, by \dots).
For each \(\phi\) we explicitly construct the corresponding fiber functor / indecomposable module category, and we
identify the associated Lagrangian (condensable) algebra \(\mathcal{A}_\phi\) inside
\(\Zc(\cRep(G)\boxtimes \Vecc_G)\simeq D(G^2)\), together with the pairing rule in Eq.~(\ref{pair}).

We interpret \(\mathcal{A}_\phi\) as the gapped domain wall \(\mathcal{B}_\phi\) in the quantum double model, and derive the anyon
transmission rules across the wall:
\([g]\mapsto[\phi(\bar g)]\) for fluxes and \(\pi\mapsto \phi^*\bar\pi\) for charges (with the appropriate
charge--flux pairing dictated by Eq.~(\ref{pair})).
These intrinsically mixed phases therefore realize domain walls whose action is invisible upon restricting to
either the pure charge (\(\Rep(G)\)) or pure flux (\(\Vecc_G\)) symmetry alone, but is fully captured by the mixed
\(\cRep(G)\boxtimes \Vecc_G\) structure.

Our strategy is to connect the abstract classification data to an explicit microscopic realization and to
directly computable observables, while keeping the categorical dictionary
\begin{equation*}
\mathcal{C}\text{-SPT} \quad \Longleftrightarrow \quad
\mathcal{C}\text{-module over } \Vecc \quad \Longleftrightarrow \quad
\text{fiber functor } \mathcal{C} \to \Vecc .
\end{equation*}
Concretely, we start from a modified quantum double construction equipped with a gapped domain wall \(\mathcal{B}_\phi\)
and compatible smooth/rough boundaries, and then perform a controlled reduction to a one-dimensional chain,
which yields a (possibly twisted) group-based cluster state.
In this reduction, closed ribbon operators descend to on-site/non-on-site symmetry operators on the chain:
one family implements the \(G\) sector (flux-type operations), while the other realizes the \(\Rep(G)\) sector
(charge-type operations), and their mutual commutation/attachment relations encode the same endomorphism data
\(\phi\) that labels our intrinsically mixed phases.
On the mathematical side we construct the corresponding module categories and match them to condensable
algebras in \(\Zc(\mathcal C)\), leveraging standard results on fusion categories and their module categories
\cite{ENO2002}.

The paper is organized as follows. 
In Sec.~\ref{Construction of the models} we construct the corresponding \(1+1\)d lattice realizations by contracting a modified Kitaev quantum
double model with a \(\mathcal B_\phi\) domain wall and smooth/rough boundaries, obtaining (possibly twisted)
group-based cluster states and their symmetry operators (closed ribbons) implementing the \(G\) and \(\Rep(G)\)
actions.
In Sec.~\ref{Characterize SPT by condensable algebras} we characterize each phase by a condensable/Lagrangian algebra \(\mathcal A_\phi\) in the bulk
\(\Zc(\mathcal H)\simeq \mathcal D_G^2\) and derive the pairing rule Eq.~(\ref{pair}).
Finally, in Sec.~\ref{Interpretation in physics} we give the SymTFT interpretation and identify \(\mathcal A_\phi\) with the
physical domain wall \(\mathcal B_\phi\) in the quantum double model, including the induced anyon transmission
rules across the wall. We conclude with a brief discussion and outlook; additional computations are collected in
the Appendix.

\smallskip
\noindent\textbf{Notation.}
Throughout, \(G\) is a finite group and \(\Vecc_G\) denotes the fusion category of finite-dimensional \(G\)-graded
complex vector spaces (with \(\Vecc\) the category of finite-dimensional complex vector spaces).
We write \(\Rep(G)\) for finite-dimensional complex representations, and \(\cRep(G)\) for the corresponding
fusion category when viewed as a symmetry input.
The Drinfeld center is denoted by \(\Zc(\mathcal C)\).
We use \([g]\) for the simple flux anyon labeled by a (conjugacy class of) group element \(g\), and \(\pi\) for a
charge anyon labeled by an irreducible representation; a bar indicates inversion/dualization (e.g. \(\bar g=g^{-1}\),
\(\bar\pi\) the dual representation), and \(\phi^*\) denotes pullback of representations along \(\phi\).
All categories are assumed \(\Bbb C\)-linear, semisimple, and unitary where appropriate. We will refer to $\Vecc_{G\times G}$, $\cRep(G) \boxtimes \Vecc_G$ and $\mathcal{D}_G^2$ as $\mathcal{G}$, $\mathcal{H}$ and $\mathcal{D}$ respectively in following context.

\section{Construction of the models}\label{Construction of the models}
In this section, we present a lattice realization of the classified SPT phases based on Kitaev's quantum double model.

\subsection{Review of Kitaev's quantum double}
We consider the \((2+1)\)-dimensional Kitaev quantum double model~\cite{KITAEV20032} defined on a honeycomb lattice. Given a finite group \(G\), a local Hilbert space \(\mathcal{H}_G=\mathbf{span}\{|g\rangle\}_{g\in G}\) is assigned to each oriented edge; reversing the orientation corresponds to inverting the group element. We represent a physical edge state by an arrow together with a group element \(g\).

For each vertex \(\alpha\), the vertex operator is defined as
\begin{equation}
    \hat{V}_\alpha
    \left[\vcenter{\hbox{
    \begin{tikzpicture}[scale=0.5]
        \node[below right] at (0,0) {$\alpha$};
        \draw[thick, decoration = {markings, mark=at position 0.6 with {\arrow[scale=1]{stealth}; \node[right]{$g_1$};}}, postaction=decorate](0,0) -- (0,-2);
        \draw[thick, decoration = {markings, mark=at position 0.6 with {\arrow[scale=1]{stealth}; \node[below]{$g_2$};}}, postaction=decorate](0,0) -- ({-sqrt(3)},1);
        \draw[thick, decoration = {markings, mark=at position 0.6 with {\arrow[scale=1]{stealth}; \node[below]{$g_3$};}}, postaction=decorate](0,0) -- ({sqrt(3)},1);
    \end{tikzpicture}
    }}\right]
    \coloneq \delta_{e,g_3 g_2 g_1}
    \left[\vcenter{\hbox{
    \begin{tikzpicture}[scale=0.5]
        \node[below right] at (0,0) {$\alpha$};
        \draw[thick, decoration = {markings, mark=at position 0.6 with {\arrow[scale=1]{stealth}; \node[right]{$g_1$};}}, postaction=decorate](0,0) -- (0,-2);
        \draw[thick, decoration = {markings, mark=at position 0.6 with {\arrow[scale=1]{stealth}; \node[below]{$g_2$};}}, postaction=decorate](0,0) -- ({-sqrt(3)},1);
        \draw[thick, decoration = {markings, mark=at position 0.6 with {\arrow[scale=1]{stealth}; \node[below]{$g_3$};}}, postaction=decorate](0,0) -- ({sqrt(3)},1);
    \end{tikzpicture}
    }}\right].
\end{equation}
For each plaquette \(\beta\), the plaquette operator associated with a group element \(h\) is defined by
\begin{equation}
    \hat{P}_\beta(h)
    \left[\vcenter{\hbox{
    \begin{tikzpicture}[x=0.75pt,y=0.75pt,yscale=-0.6,xscale=0.6]
    %uncomment if require: \path (0,393); %set diagram left start at 0, and has height of 393
    
    %Shape: Regular Polygon [id:dp819861996763323] 
    \draw   (379.09,206.87) -- (355.57,247.62) -- (308.52,247.62) -- (285,206.87) -- (308.52,166.13) -- (355.57,166.13) -- cycle ;
    %Straight Lines [id:da14751301123472227] 
    \draw  [dash pattern={on 4.5pt off 4.5pt}]  (379.09,206.87) -- (402.62,206.87) ;
    %Straight Lines [id:da27340360817593323] 
    \draw  [dash pattern={on 4.5pt off 4.5pt}]  (355.57,247.62) -- (367.33,267.99) ;
    %Straight Lines [id:da5967876878449745] 
    \draw  [dash pattern={on 4.5pt off 4.5pt}]  (285,206.87) -- (261.48,206.87) ;
    %Straight Lines [id:da7765799891045057] 
    \draw  [dash pattern={on 4.5pt off 4.5pt}]  (296.76,267.99) -- (308.52,247.62) ;
    %Straight Lines [id:da8086896618124784] 
    \draw  [dash pattern={on 4.5pt off 4.5pt}]  (355.57,166.13) -- (367.33,145.76) ;
    %Straight Lines [id:da8757177488381989] 
    \draw  [dash pattern={on 4.5pt off 4.5pt}]  (296.76,145.76) -- (308.52,166.13) ;
    %Straight Lines [id:da3235483756239407] 
    \draw [line width=0.75]    (308.52,166.13) -- (285,206.87) ;
    \draw [shift={(300.01,180.87)}, rotate = 120] [fill={rgb, 255:red, 0; green, 0; blue, 0 }  ][line width=0.08]  [draw opacity=0] (10.72,-5.15) -- (0,0) -- (10.72,5.15) -- (7.12,0) -- cycle    ;
    %Straight Lines [id:da38419172942147295] 
    \draw [line width=0.75]    (308.52,166.13) -- (355.57,166.13) ;
    \draw [shift={(337.05,166.13)}, rotate = 180] [fill={rgb, 255:red, 0; green, 0; blue, 0 }  ][line width=0.08]  [draw opacity=0] (10.72,-5.15) -- (0,0) -- (10.72,5.15) -- (7.12,0) -- cycle    ;
    %Straight Lines [id:da031658318164157984] 
    \draw [line width=0.75]    (355.57,166.13) -- (379.09,206.87) ;
    \draw [shift={(369.83,190.83)}, rotate = 240] [fill={rgb, 255:red, 0; green, 0; blue, 0 }  ][line width=0.08]  [draw opacity=0] (10.72,-5.15) -- (0,0) -- (10.72,5.15) -- (7.12,0) -- cycle    ;
    %Straight Lines [id:da22878233381549018] 
    \draw [line width=0.75]    (379.09,206.87) -- (355.57,247.62) ;
    \draw [shift={(364.83,231.57)}, rotate = 300] [fill={rgb, 255:red, 0; green, 0; blue, 0 }  ][line width=0.08]  [draw opacity=0] (10.72,-5.15) -- (0,0) -- (10.72,5.15) -- (7.12,0) -- cycle    ;
    %Straight Lines [id:da7244634088088912] 
    \draw    (355.57,247.62) -- (308.52,247.62) ;
    \draw [shift={(327.05,247.62)}, rotate = 360] [fill={rgb, 255:red, 0; green, 0; blue, 0 }  ][line width=0.08]  [draw opacity=0] (10.72,-5.15) -- (0,0) -- (10.72,5.15) -- (7.12,0) -- cycle    ;
    %Straight Lines [id:da365022882159734] 
    \draw    (285,206.87) -- (308.52,247.62) ;
    \draw [shift={(293.51,221.62)}, rotate = 60] [fill={rgb, 255:red, 0; green, 0; blue, 0 }  ][line width=0.08]  [draw opacity=0] (10.72,-5.15) -- (0,0) -- (10.72,5.15) -- (7.12,0) -- cycle    ;

    % Text Node
    \draw (285.34,179.91) node [anchor=south] [inner sep=0.75pt]  [rotate=-300] [align=left] {$\displaystyle g_{1}$};
    % Text Node
    \draw (332.05,156.13) node [anchor=south] [inner sep=0.75pt]   [align=left] {$\displaystyle g_{2}$};
    % Text Node
    \draw (332.05,250.62) node [anchor=north] [inner sep=0.75pt]   [align=left] {$\displaystyle g_{5}$};
    % Text Node
    \draw (377.93,179.91) node [anchor=south] [inner sep=0.75pt]  [rotate=-60] [align=left] {$\displaystyle g_{3}$};
    % Text Node
    \draw (369.93,228.74) node [anchor=north] [inner sep=0.75pt]  [rotate=-300] [align=left] {$\displaystyle g_{4}$};
    % Text Node
    \draw (292.79,228.74) node [anchor=north] [inner sep=0.75pt]  [rotate=-60] [align=left] {$\displaystyle g_{6}$};
    % Text Node
    \draw (327.05,197.87) node [anchor=north west][inner sep=0.75pt]   [align=left] {$\displaystyle \beta $};

    \end{tikzpicture}
    }}\right]
    \coloneq \left[\vcenter{\hbox{
    \begin{tikzpicture}[x=0.75pt,y=0.75pt,yscale=-0.6,xscale=0.6]
    %uncomment if require: \path (0,393); %set diagram left start at 0, and has height of 393
    
    %Shape: Regular Polygon [id:dp819861996763323] 
    \draw   (378.09,206.87) -- (354.57,247.62) -- (307.52,247.62) -- (284,206.87) -- (307.52,166.13) -- (354.57,166.13) -- cycle ;
    %Straight Lines [id:da14751301123472227] 
    \draw  [dash pattern={on 4.5pt off 4.5pt}]  (378.09,206.87) -- (401.62,206.87) ;
    %Straight Lines [id:da27340360817593323] 
    \draw  [dash pattern={on 4.5pt off 4.5pt}]  (354.57,247.62) -- (366.33,267.99) ;
    %Straight Lines [id:da5967876878449745] 
    \draw  [dash pattern={on 4.5pt off 4.5pt}]  (284,206.87) -- (260.48,206.87) ;
    %Straight Lines [id:da7765799891045057] 
    \draw  [dash pattern={on 4.5pt off 4.5pt}]  (295.76,267.99) -- (307.52,247.62) ;
    %Straight Lines [id:da8086896618124784] 
    \draw  [dash pattern={on 4.5pt off 4.5pt}]  (354.57,166.13) -- (366.33,145.76) ;
    %Straight Lines [id:da8757177488381989] 
    \draw  [dash pattern={on 4.5pt off 4.5pt}]  (295.76,145.76) -- (307.52,166.13) ;
    %Straight Lines [id:da3235483756239407] 
    \draw [line width=0.75]    (307.52,166.13) -- (284,206.87) ;
    \draw [shift={(299.01,180.87)}, rotate = 120] [fill={rgb, 255:red, 0; green, 0; blue, 0 }  ][line width=0.08]  [draw opacity=0] (10.72,-5.15) -- (0,0) -- (10.72,5.15) -- (7.12,0) -- cycle    ;
    %Straight Lines [id:da38419172942147295] 
    \draw [line width=0.75]    (307.52,166.13) -- (354.57,166.13) ;
    \draw [shift={(336.05,166.13)}, rotate = 180] [fill={rgb, 255:red, 0; green, 0; blue, 0 }  ][line width=0.08]  [draw opacity=0] (10.72,-5.15) -- (0,0) -- (10.72,5.15) -- (7.12,0) -- cycle    ;
    %Straight Lines [id:da031658318164157984] 
    \draw [line width=0.75]    (354.57,166.13) -- (378.09,206.87) ;
    \draw [shift={(368.83,190.83)}, rotate = 240] [fill={rgb, 255:red, 0; green, 0; blue, 0 }  ][line width=0.08]  [draw opacity=0] (10.72,-5.15) -- (0,0) -- (10.72,5.15) -- (7.12,0) -- cycle    ;
    %Straight Lines [id:da22878233381549018] 
    \draw [line width=0.75]    (378.09,206.87) -- (354.57,247.62) ;
    \draw [shift={(363.83,231.57)}, rotate = 300] [fill={rgb, 255:red, 0; green, 0; blue, 0 }  ][line width=0.08]  [draw opacity=0] (10.72,-5.15) -- (0,0) -- (10.72,5.15) -- (7.12,0) -- cycle    ;
    %Straight Lines [id:da7244634088088912] 
    \draw    (354.57,247.62) -- (307.52,247.62) ;
    \draw [shift={(326.05,247.62)}, rotate = 360] [fill={rgb, 255:red, 0; green, 0; blue, 0 }  ][line width=0.08]  [draw opacity=0] (10.72,-5.15) -- (0,0) -- (10.72,5.15) -- (7.12,0) -- cycle    ;
    %Straight Lines [id:da365022882159734] 
    \draw    (284,206.87) -- (307.52,247.62) ;
    \draw [shift={(292.51,221.62)}, rotate = 60] [fill={rgb, 255:red, 0; green, 0; blue, 0 }  ][line width=0.08]  [draw opacity=0] (10.72,-5.15) -- (0,0) -- (10.72,5.15) -- (7.12,0) -- cycle    ;

    % Text Node
    \draw (284.34,179.91) node [anchor=south] [inner sep=0.75pt]  [rotate=-300] [align=left] {$\displaystyle hg_{1}$};
    % Text Node
    \draw (331.05,156.13) node [anchor=south] [inner sep=0.75pt]   [align=left] {$\displaystyle hg_{2}$};
    % Text Node
    \draw (331.05,250.62) node [anchor=north] [inner sep=0.75pt]   [align=left] {$\displaystyle hg_{5}$};
    % Text Node
    \draw (376.93,179.91) node [anchor=south] [inner sep=0.75pt]  [rotate=-60] [align=left] {$\displaystyle hg_{3}$};
    % Text Node
    \draw (368.93,228.74) node [anchor=north] [inner sep=0.75pt]  [rotate=-300] [align=left] {$\displaystyle hg_{4}$};
    % Text Node
    \draw (291.79,228.74) node [anchor=north] [inner sep=0.75pt]  [rotate=-60] [align=left] {$\displaystyle hg_{6}$};
    % Text Node
    \draw (327.05,197.87) node [anchor=north west][inner sep=0.75pt]   [align=left] {$\displaystyle \beta $};

    \end{tikzpicture}

    }}\right],
\end{equation}
or, equivalently, \(\hat{P}_\beta(h)=\prod_{i\in \beta}L_h^{(i)}\), where \(L(R)_h^{(i)}\) denotes the left (right) action of \(h\) on edge \(i\), which is equal to the right (left) action of \(\bar{h}\) when the orientation of edge \(i\) is reversed. The total plaquette operator is \(\hat{P}_\beta=\frac{1}{|G|}\sum_{h\in G} \hat{P}_\beta(h)\).

The Hamiltonian of Kitaev's quantum double model is
\begin{equation}
    H=-\sum_{\text{vertex }\alpha}\hat{V}_\alpha - \sum_{\text{plaquette }\beta}\hat{P}_\beta.
\end{equation}
Note that \(\hat{V}_\alpha\) and \(\hat{P}_\beta\) are projectors and mutually commute for all \(\alpha\) and \(\beta\). Consequently, the ground-state subspace consists of the common \(+1\) eigenspaces of all \(\hat{V}_\alpha\) and \(\hat{P}_\beta\).

A quasiparticle excitation is specified by a conjugacy class \([g]\) together with an irreducible representation \(\pi\) of the centralizer of a chosen representative \(g\) of that conjugacy class. In addition, there are non-topological (local) degrees of freedom: one may label them by an element \(pg\bar{p}\in[g]\) and a basis vector \(\hat{e}_i\) of the representation space, and denote the corresponding excitation by \(([g],\pi)_{pi}\) (often suppressing the non-topological index when it is not essential). Quasiparticles are created and transported by ribbon operators. 

A general ribbon operator, labeled by a pair of group elements \(g\) and \(h\), is defined on a complete path: namely, a path \(P\) on the direct lattice (with edge indices \(i\in I\)) together with a path \(P^*\) on the dual lattice (with dual-edge indices \(i^*\in I^*\)) such that \(P^*\) passes through every plaquette immediately to the right of \(P\). The corresponding ribbon operator takes the form
\begin{equation}
    F^{h,g}(P)= \delta_{e,\prod_{i^* \in I^*}g_{i^*}}\cdot \left(\prod_{i^*\in I^*,0}\prod_{i\in I_{i^*}}L_{(g_{i^*}...g_{1^*})h(\bar{g}_{1^*}...\bar{g}_{i^*})}^{(i)}\right),
\end{equation}
where \(I_{i^*}\) denotes the set of edges of \(P\) belonging to the plaquette adjacent to the dual edges \(i^*\) and \(i^*+1\). For example:

\begin{equation}
    \begin{aligned}
    F^{h,g}(\text{red path})\left[
    \vcenter{\hbox{
    \begin{tikzpicture}[x=0.75pt,y=0.75pt,yscale=-1,xscale=1]
%uncomment if require: \path (0,236); %set diagram left start at 0, and has height of 236

%Shape: Regular Polygon [id:dp5320106013377258] 
\draw   (211,148) -- (198.5,169.65) -- (173.5,169.65) -- (161,148) -- (173.5,126.35) -- (198.5,126.35) -- cycle ;
%Shape: Regular Polygon [id:dp962122683736305] 
\draw   (248.5,126.35) -- (236,148) -- (211,148) -- (198.5,126.35) -- (211,104.7) -- (236,104.7) -- cycle ;
%Shape: Regular Polygon [id:dp9319659872017714] 
\draw   (248.5,169.65) -- (236,191.3) -- (211,191.3) -- (198.5,169.65) -- (211,148) -- (236,148) -- cycle ;
%Shape: Regular Polygon [id:dp5291373682434122] 
\draw   (286,148) -- (273.5,169.65) -- (248.5,169.65) -- (236,148) -- (248.5,126.35) -- (273.5,126.35) -- cycle ;
%Straight Lines [id:da9503350871546868] 
\draw    (248.5,169.65) -- (236,191.3) ;
%Straight Lines [id:da0941719712431155] 
\draw    (173.5,126.35) -- (161,148) ;
%Straight Lines [id:da2502657649022575] 
\draw [color={rgb, 255:red, 208; green, 2; blue, 27 }  ,draw opacity=1 ]   (198.5,126.35) -- (211,148) ;
\draw [shift={(206.55,140.29)}, rotate = 240] [color={rgb, 255:red, 208; green, 2; blue, 27 }  ,draw opacity=1 ][line width=0.75]    (6.56,-1.97) .. controls (4.17,-0.84) and (1.99,-0.18) .. (0,0) .. controls (1.99,0.18) and (4.17,0.84) .. (6.56,1.97)   ;
%Straight Lines [id:da6829196188804417] 
\draw    (211,148) -- (236,148) ;
%Straight Lines [id:da37836781459351476] 
\draw [color={rgb, 255:red, 208; green, 2; blue, 27 }  ,draw opacity=1 ]   (211,148) -- (236,148) ;
\draw [shift={(227.1,148)}, rotate = 180] [color={rgb, 255:red, 208; green, 2; blue, 27 }  ,draw opacity=1 ][line width=0.75]    (6.56,-1.97) .. controls (4.17,-0.84) and (1.99,-0.18) .. (0,0) .. controls (1.99,0.18) and (4.17,0.84) .. (6.56,1.97)   ;
%Straight Lines [id:da5348315352086426] 
\draw [color={rgb, 255:red, 208; green, 2; blue, 27 }  ,draw opacity=1 ]   (236,148) -- (248.5,126.35) ;
\draw [shift={(244.05,134.06)}, rotate = 120] [color={rgb, 255:red, 208; green, 2; blue, 27 }  ,draw opacity=1 ][line width=0.75]    (6.56,-1.97) .. controls (4.17,-0.84) and (1.99,-0.18) .. (0,0) .. controls (1.99,0.18) and (4.17,0.84) .. (6.56,1.97)   ;
%Shape: Regular Polygon [id:dp8213690380309264] 
\draw   (211,104.7) -- (198.5,126.35) -- (173.5,126.35) -- (161,104.7) -- (173.5,83.05) -- (198.5,83.05) -- cycle ;
%Straight Lines [id:da0299135645053763] 
\draw [color={rgb, 255:red, 208; green, 2; blue, 27 }  ,draw opacity=1 ]   (173.5,126.35) -- (198.5,126.35) ;
\draw [shift={(189.6,126.35)}, rotate = 180] [color={rgb, 255:red, 208; green, 2; blue, 27 }  ,draw opacity=1 ][line width=0.75]    (6.56,-1.97) .. controls (4.17,-0.84) and (1.99,-0.18) .. (0,0) .. controls (1.99,0.18) and (4.17,0.84) .. (6.56,1.97)   ;
%Shape: Regular Polygon [id:dp4004543343446151] 
\draw   (286,104.7) -- (273.5,126.35) -- (248.5,126.35) -- (236,104.7) -- (248.5,83.05) -- (273.5,83.05) -- cycle ;
%Straight Lines [id:da5691875433466684] 
\draw [color={rgb, 255:red, 208; green, 2; blue, 27 }  ,draw opacity=1 ]   (248.5,126.35) -- (273.5,126.35) ;
\draw [shift={(264.6,126.35)}, rotate = 180] [color={rgb, 255:red, 208; green, 2; blue, 27 }  ,draw opacity=1 ][line width=0.75]    (6.56,-1.97) .. controls (4.17,-0.84) and (1.99,-0.18) .. (0,0) .. controls (1.99,0.18) and (4.17,0.84) .. (6.56,1.97)   ;
%Straight Lines [id:da2939005046364912] 
\draw [color={rgb, 255:red, 74; green, 144; blue, 226 }  ,draw opacity=1 ]   (211,148) -- (198.5,169.65) ;
\draw [shift={(207.05,154.84)}, rotate = 120] [color={rgb, 255:red, 74; green, 144; blue, 226 }  ,draw opacity=1 ][line width=0.75]    (6.56,-1.97) .. controls (4.17,-0.84) and (1.99,-0.18) .. (0,0) .. controls (1.99,0.18) and (4.17,0.84) .. (6.56,1.97)   ;
%Straight Lines [id:da97100924228727] 
\draw [color={rgb, 255:red, 74; green, 144; blue, 226 }  ,draw opacity=1 ]   (236,148) -- (248.5,169.65) ;
\draw [shift={(239.95,154.84)}, rotate = 60] [color={rgb, 255:red, 74; green, 144; blue, 226 }  ,draw opacity=1 ][line width=0.75]    (6.56,-1.97) .. controls (4.17,-0.84) and (1.99,-0.18) .. (0,0) .. controls (1.99,0.18) and (4.17,0.84) .. (6.56,1.97)   ;
%Straight Lines [id:da1361694179314089] 
\draw [color={rgb, 255:red, 65; green, 117; blue, 5 }  ,draw opacity=1 ][line width=1.5]    (186,148) -- (204.75,158.83) ;
%Shape: Arc [id:dp6282651679959022] 
\draw  [draw opacity=0][line width=1.5]  (242.26,158.83) .. controls (236.74,162.02) and (230.33,163.84) .. (223.5,163.84) .. controls (216.67,163.84) and (210.27,162.02) .. (204.75,158.83) -- (223.5,126.34) -- cycle ; \draw [color={rgb, 255:red, 65; green, 117; blue, 5 }  ,draw opacity=1 ][line width=1.5]    (238.67,160.65) .. controls (234.03,162.7) and (228.9,163.84) .. (223.5,163.84) .. controls (216.67,163.84) and (210.27,162.02) .. (204.75,158.83) ;  \draw [shift={(242.26,158.83)}, rotate = 156.12] [fill={rgb, 255:red, 65; green, 117; blue, 5 }  ,fill opacity=1 ][line width=0.08]  [draw opacity=0] (8.75,-4.2) -- (0,0) -- (8.75,4.2) -- (5.81,0) -- cycle    ;
%Straight Lines [id:da2641847631843558] 
\draw [color={rgb, 255:red, 65; green, 117; blue, 5 }  ,draw opacity=1 ][line width=1.5]    (242.26,158.83) -- (261,148) ;

% Text Node
\draw (186,115.52) node  [font=\tiny] [align=left] {$\displaystyle {\textstyle \textcolor[rgb]{0.82,0.01,0.11}{y}\textcolor[rgb]{0.82,0.01,0.11}{_{1}}}$};
% Text Node
\draw (202.75,137.17) node [anchor=east] [inner sep=0.75pt]  [font=\tiny] [align=left] {$\displaystyle {\textstyle \textcolor[rgb]{0.82,0.01,0.11}{y}\textcolor[rgb]{0.82,0.01,0.11}{_{2}}}$};
% Text Node
\draw (223.5,151) node [anchor=north] [inner sep=0.75pt]  [font=\tiny] [align=left] {$\displaystyle \textcolor[rgb]{0.82,0.01,0.11}{y}\textcolor[rgb]{0.82,0.01,0.11}{_{3}}$};
% Text Node
\draw (240.25,134.17) node [anchor=south east] [inner sep=0.75pt]  [font=\tiny] [align=left] {$\displaystyle {\textstyle \textcolor[rgb]{0.82,0.01,0.11}{y}\textcolor[rgb]{0.82,0.01,0.11}{_{4}}}$};
% Text Node
\draw (261.5,118.52) node [anchor=south] [inner sep=0.75pt]  [font=\tiny] [align=left] {$\displaystyle {\textstyle \textcolor[rgb]{0.82,0.01,0.11}{y}\textcolor[rgb]{0.82,0.01,0.11}{_{5}}}$};
% Text Node
\draw (209,148) node [anchor=east] [inner sep=0.75pt]  [font=\tiny] [align=left] {$\displaystyle {\textstyle \textcolor[rgb]{0.29,0.56,0.89}{x}\textcolor[rgb]{0.29,0.56,0.89}{_{1}}}$};
% Text Node
\draw (238,148) node [anchor=west] [inner sep=0.75pt]  [font=\tiny] [align=left] {$\displaystyle \textcolor[rgb]{0.29,0.56,0.89}{x}\textcolor[rgb]{0.29,0.56,0.89}{_{2}}$};
% Text Node
\draw (186,145) node [anchor=south] [inner sep=0.75pt]  [font=\tiny] [align=left] {$\displaystyle g$};
% Text Node
\draw (186,151) node [anchor=north] [inner sep=0.75pt]  [font=\tiny] [align=left] {$\displaystyle \textcolor[rgb]{0.25,0.46,0.02}{{\textstyle h}}$};
% Text Node
\draw (223.5,174.15) node  [font=\tiny] [align=left] {$\displaystyle {\textstyle \textcolor[rgb]{0.25,0.46,0.02}{i}\textcolor[rgb]{0.25,0.46,0.02}{_{x_{1}}}\textcolor[rgb]{0.25,0.46,0.02}{h}}$};
% Text Node
\draw (261,145) node [anchor=south] [inner sep=0.75pt]  [font=\tiny] [align=left] {$\displaystyle {\textstyle \textcolor[rgb]{0.25,0.46,0.02}{i}\textcolor[rgb]{0.25,0.46,0.02}{_{x_{2} x_{1}}}\textcolor[rgb]{0.25,0.46,0.02}{h}}$};

\end{tikzpicture}
}}\right]
    =
    \delta_{g,x_2x_1}
    \left[\vcenter{\hbox{
    \begin{tikzpicture}[x=0.75pt,y=0.75pt,yscale=-1,xscale=1]
%uncomment if require: \path (0,326); %set diagram left start at 0, and has height of 326

%Shape: Polygon [id:dp10359253901916177] 
\draw   (209.78,145.1) -- (197.59,166.77) -- (173.2,166.77) -- (161,145.1) -- (173.2,123.42) -- (197.59,123.42) -- cycle ;
%Shape: Polygon [id:dp009181870854718444] 
\draw   (246.37,123.42) -- (234.18,145.1) -- (209.78,145.1) -- (197.59,123.42) -- (209.78,101.75) -- (234.18,101.75) -- cycle ;
%Shape: Polygon [id:dp656085460988033] 
\draw   (246.37,166.77) -- (234.18,188.45) -- (209.78,188.45) -- (197.59,166.77) -- (209.78,145.1) -- (234.18,145.1) -- cycle ;
%Shape: Polygon [id:dp9702012976409193] 
\draw   (282.96,145.1) -- (270.76,166.77) -- (246.37,166.77) -- (234.18,145.1) -- (246.37,123.42) -- (270.76,123.42) -- cycle ;
%Straight Lines [id:da4440214848170885] 
\draw    (246.37,166.77) -- (234.18,188.45) ;
%Straight Lines [id:da27748439413889237] 
\draw    (173.2,123.42) -- (161,145.1) ;
%Straight Lines [id:da8566573774145165] 
\draw [color={rgb, 255:red, 208; green, 2; blue, 27 }  ,draw opacity=1 ][line width=0.75]    (197.59,123.42) -- (209.78,145.1) ;
\draw [shift={(205.45,137.4)}, rotate = 240.64] [color={rgb, 255:red, 208; green, 2; blue, 27 }  ,draw opacity=1 ][line width=0.75]    (6.56,-1.97) .. controls (4.17,-0.84) and (1.99,-0.18) .. (0,0) .. controls (1.99,0.18) and (4.17,0.84) .. (6.56,1.97)   ;
%Straight Lines [id:da8630954328533478] 
\draw    (209.78,145.1) -- (234.18,145.1) ;
%Straight Lines [id:da877037156016211] 
\draw [color={rgb, 255:red, 208; green, 2; blue, 27 }  ,draw opacity=1 ][line width=0.75]    (209.78,145.1) -- (234.18,145.1) ;
\draw [shift={(225.58,145.1)}, rotate = 180] [color={rgb, 255:red, 208; green, 2; blue, 27 }  ,draw opacity=1 ][line width=0.75]    (6.56,-1.97) .. controls (4.17,-0.84) and (1.99,-0.18) .. (0,0) .. controls (1.99,0.18) and (4.17,0.84) .. (6.56,1.97)   ;
%Straight Lines [id:da8232930818141466] 
\draw [color={rgb, 255:red, 208; green, 2; blue, 27 }  ,draw opacity=1 ][line width=0.75]    (234.18,145.1) -- (246.37,123.42) ;
\draw [shift={(242.04,131.12)}, rotate = 119.36] [color={rgb, 255:red, 208; green, 2; blue, 27 }  ,draw opacity=1 ][line width=0.75]    (6.56,-1.97) .. controls (4.17,-0.84) and (1.99,-0.18) .. (0,0) .. controls (1.99,0.18) and (4.17,0.84) .. (6.56,1.97)   ;
%Shape: Polygon [id:dp5092513189165856] 
\draw   (209.78,101.75) -- (197.59,123.42) -- (173.2,123.42) -- (161,101.75) -- (173.2,80.07) -- (197.59,80.07) -- cycle ;
%Straight Lines [id:da6212476266503252] 
\draw [color={rgb, 255:red, 208; green, 2; blue, 27 }  ,draw opacity=1 ][line width=0.75]    (173.2,123.42) -- (197.59,123.42) ;
\draw [shift={(188.99,123.42)}, rotate = 180] [color={rgb, 255:red, 208; green, 2; blue, 27 }  ,draw opacity=1 ][line width=0.75]    (6.56,-1.97) .. controls (4.17,-0.84) and (1.99,-0.18) .. (0,0) .. controls (1.99,0.18) and (4.17,0.84) .. (6.56,1.97)   ;
%Shape: Polygon [id:dp7153842050888325] 
\draw  [color={rgb, 255:red, 155; green, 155; blue, 155 }  ,draw opacity=1 ] (282.96,101.75) -- (270.76,123.42) -- (246.37,123.42) -- (234.18,101.75) -- (246.37,80.07) -- (270.76,80.07) -- cycle ;
%Straight Lines [id:da5262777426165263] 
\draw [color={rgb, 255:red, 208; green, 2; blue, 27 }  ,draw opacity=1 ][line width=0.75]    (246.37,123.42) -- (270.76,123.42) ;
\draw [shift={(262.17,123.42)}, rotate = 180] [color={rgb, 255:red, 208; green, 2; blue, 27 }  ,draw opacity=1 ][line width=0.75]    (6.56,-1.97) .. controls (4.17,-0.84) and (1.99,-0.18) .. (0,0) .. controls (1.99,0.18) and (4.17,0.84) .. (6.56,1.97)   ;
%Straight Lines [id:da6686954026641829] 
\draw [color={rgb, 255:red, 74; green, 144; blue, 226 }  ,draw opacity=1 ][line width=0.75]    (209.78,145.1) -- (197.59,166.77) ;
\draw [shift={(205.94,151.93)}, rotate = 119.36] [color={rgb, 255:red, 74; green, 144; blue, 226 }  ,draw opacity=1 ][line width=0.75]    (6.56,-1.97) .. controls (4.17,-0.84) and (1.99,-0.18) .. (0,0) .. controls (1.99,0.18) and (4.17,0.84) .. (6.56,1.97)   ;
%Straight Lines [id:da6845739497031754] 
\draw [color={rgb, 255:red, 74; green, 144; blue, 226 }  ,draw opacity=1 ][line width=0.75]    (234.18,145.1) -- (246.37,166.77) ;
\draw [shift={(238.02,151.93)}, rotate = 60.64] [color={rgb, 255:red, 74; green, 144; blue, 226 }  ,draw opacity=1 ][line width=0.75]    (6.56,-1.97) .. controls (4.17,-0.84) and (1.99,-0.18) .. (0,0) .. controls (1.99,0.18) and (4.17,0.84) .. (6.56,1.97)   ;

% Text Node
\draw (185.39,116.26) node  [font=\tiny] [align=left] {$\displaystyle \textcolor[rgb]{0.82,0.01,0.11}{h\cdot y}\textcolor[rgb]{0.82,0.01,0.11}{_{1}}$};
% Text Node
\draw (205.69,131.26) node [anchor=south west] [inner sep=0.75pt]  [font=\tiny] [align=left] {$\displaystyle \textcolor[rgb]{0.82,0.01,0.11}{h\cdotp y}\textcolor[rgb]{0.82,0.01,0.11}{_{2}}$};
% Text Node
\draw (221.98,148.1) node [anchor=north] [inner sep=0.75pt]  [font=\tiny] [align=left] {$\displaystyle \textcolor[rgb]{0.82,0.01,0.11}{i}\textcolor[rgb]{0.82,0.01,0.11}{_{x_{1}}}\textcolor[rgb]{0.82,0.01,0.11}{h\cdotp y}\textcolor[rgb]{0.82,0.01,0.11}{_{3}}$};
% Text Node
\draw (242.27,137.26) node [anchor=north west][inner sep=0.75pt]  [font=\tiny] [align=left] {$\displaystyle \textcolor[rgb]{0.82,0.01,0.11}{i}\textcolor[rgb]{0.82,0.01,0.11}{_{x_{2} x_{1}}}\textcolor[rgb]{0.82,0.01,0.11}{h\cdot y}\textcolor[rgb]{0.82,0.01,0.11}{_{4}}$};
% Text Node
\draw (258.57,120.42) node [anchor=south] [inner sep=0.75pt]  [font=\tiny] [align=left] {$\displaystyle \textcolor[rgb]{0.82,0.01,0.11}{i}\textcolor[rgb]{0.82,0.01,0.11}{_{x_{2} x_{1}}}\textcolor[rgb]{0.82,0.01,0.11}{h\cdotp y}\textcolor[rgb]{0.82,0.01,0.11}{_{5}}$};
% Text Node
\draw (207.78,145.1) node [anchor=east] [inner sep=0.75pt]  [font=\tiny] [align=left] {$\displaystyle \textcolor[rgb]{0.29,0.56,0.89}{x}\textcolor[rgb]{0.29,0.56,0.89}{_{1}}$};
% Text Node
\draw (236.18,145.1) node [anchor=west] [inner sep=0.75pt]  [font=\tiny] [align=left] {$\displaystyle \textcolor[rgb]{0.29,0.56,0.89}{x}\textcolor[rgb]{0.29,0.56,0.89}{_{2}}$};

\end{tikzpicture}
    }}
    \right]
    \end{aligned}
\end{equation}
where $i_x (h)=xh\bar{x}$. 

Simple anyons can be obtained by taking appropriate linear combinations of ribbon operators. Concretely, one defines
\begin{equation}
W_{pi,qj}^{([h],\pi)}(\text{path}) = \sum_{z\in Z(h)} \pi(z)_{ij} F^{ph\bar{p},qz\bar{p}}(\text{path}),
\end{equation}
which creates, at the terminating site of the path, a simple anyon labeled by \(([g],\pi)\), and simultaneously creates its dual anyon \(([\bar{g}],\bar{\pi})\) at the starting site. The labels \(pi\) and \(qj\) encode the internal non-topological degrees of freedom at the terminal and initial sites, respectively.

\subsection{Twisted group-based cluster state from modified quantum double}
In this section we construct a family of one-dimensional qudit chains obtained from a suitably modified quantum double model, and clarify their relation to the group-based cluster state~\cite{PhysRevX.15.011058}. Our construction starts from the defect structures of the quantum double model introduced in the previous section. the motivation for adopting this particular setup will be explained in Sec.~\ref{Interpretation in physics}.

\begin{figure}[h]
    \centering
    \begin{tikzpicture}[x=0.75pt,y=0.75pt,yscale=-1,xscale=1]
%uncomment if require: \path (0,372); %set diagram left start at 0, and has height of 372

%Shape: Regular Polygon [id:dp92917507809399] 
\draw   (308.55,152.5) -- (286.9,165) -- (265.25,152.5) -- (265.25,127.5) -- (286.9,115) -- (308.55,127.5) -- cycle ;
%Shape: Regular Polygon [id:dp3470317978007533] 
\draw   (265.25,152.5) -- (243.6,165) -- (221.95,152.5) -- (221.95,127.5) -- (243.6,115) -- (265.25,127.5) -- cycle ;
%Shape: Regular Polygon [id:dp12353221500534861] 
\draw   (221.95,152.5) -- (200.3,165) -- (178.65,152.5) -- (178.65,127.5) -- (200.3,115) -- (221.95,127.5) -- cycle ;
%Shape: Regular Polygon [id:dp09661189903868683] 
\draw   (286.9,115) -- (265.25,127.5) -- (243.6,115) -- (243.6,90) -- (265.25,77.5) -- (286.9,90) -- cycle ;
%Shape: Regular Polygon [id:dp8500374994036151] 
\draw   (243.6,115) -- (221.95,127.5) -- (200.3,115) -- (200.3,90) -- (221.95,77.5) -- (243.6,90) -- cycle ;
%Shape: Regular Polygon [id:dp41547936262434504] 
\draw   (200.3,115) -- (178.65,127.5) -- (157,115) -- (157,90) -- (178.65,77.5) -- (200.3,90) -- cycle ;
%Shape: Regular Polygon [id:dp6029560791612047] 
\draw   (178.65,152.5) -- (157,165) -- (135.35,152.5) -- (135.35,127.5) -- (157,115) -- (178.65,127.5) -- cycle ;
%Straight Lines [id:da8586895863950201] 
\draw    (157,165) -- (157,190) ;
%Straight Lines [id:da06765177738501216] 
\draw [color={rgb, 255:red, 208; green, 2; blue, 27 }  ,draw opacity=1 ][line width=0.75]    (200.3,165) -- (200.3,190) ;
\draw [shift={(200.3,172.9)}, rotate = 90] [color={rgb, 255:red, 208; green, 2; blue, 27 }  ,draw opacity=1 ][line width=0.75]    (6.56,-1.97) .. controls (4.17,-0.84) and (1.99,-0.18) .. (0,0) .. controls (1.99,0.18) and (4.17,0.84) .. (6.56,1.97)   ;
%Straight Lines [id:da024672582822805267] 
\draw    (243.6,165) -- (243.6,190) ;
%Straight Lines [id:da5831234228373972] 
\draw    (286.9,165) -- (286.9,190) ;
%Straight Lines [id:da9894237090255343] 
\draw    (200.3,165) -- (221.95,152.5) ;
%Straight Lines [id:da7911251311692319] 
\draw    (221.95,152.5) -- (243.6,165) ;
%Straight Lines [id:da987022843086015] 
\draw    (243.6,165) -- (243.6,190) ;
%Straight Lines [id:da23982517636596867] 
\draw [color={rgb, 255:red, 208; green, 2; blue, 27 }  ,draw opacity=1 ][line width=0.75]    (200.3,165) -- (221.95,152.5) ;
\draw [shift={(214.24,156.95)}, rotate = 150] [color={rgb, 255:red, 208; green, 2; blue, 27 }  ,draw opacity=1 ][line width=0.75]    (6.56,-1.97) .. controls (4.17,-0.84) and (1.99,-0.18) .. (0,0) .. controls (1.99,0.18) and (4.17,0.84) .. (6.56,1.97)   ;
%Straight Lines [id:da6781386701727067] 
\draw [color={rgb, 255:red, 208; green, 2; blue, 27 }  ,draw opacity=1 ][line width=0.75]    (221.95,152.5) -- (243.6,165) ;
\draw [shift={(235.89,160.55)}, rotate = 210] [color={rgb, 255:red, 208; green, 2; blue, 27 }  ,draw opacity=1 ][line width=0.75]    (6.56,-1.97) .. controls (4.17,-0.84) and (1.99,-0.18) .. (0,0) .. controls (1.99,0.18) and (4.17,0.84) .. (6.56,1.97)   ;
%Straight Lines [id:da27507241314736586] 
\draw [color={rgb, 255:red, 208; green, 2; blue, 27 }  ,draw opacity=1 ][line width=0.75]    (243.6,165) -- (243.6,190) ;
\draw [shift={(243.6,181.1)}, rotate = 270] [color={rgb, 255:red, 208; green, 2; blue, 27 }  ,draw opacity=1 ][line width=0.75]    (6.56,-1.97) .. controls (4.17,-0.84) and (1.99,-0.18) .. (0,0) .. controls (1.99,0.18) and (4.17,0.84) .. (6.56,1.97)   ;
%Straight Lines [id:da4794035227550426] 
\draw [color={rgb, 255:red, 208; green, 2; blue, 27 }  ,draw opacity=1 ][line width=0.75]    (221.95,77.5) -- (200.3,90) ;
\draw [shift={(215.11,81.45)}, rotate = 150] [color={rgb, 255:red, 208; green, 2; blue, 27 }  ,draw opacity=1 ][line width=0.75]    (6.56,-1.97) .. controls (4.17,-0.84) and (1.99,-0.18) .. (0,0) .. controls (1.99,0.18) and (4.17,0.84) .. (6.56,1.97)   ;
%Straight Lines [id:da02909933101781137] 
\draw [color={rgb, 255:red, 208; green, 2; blue, 27 }  ,draw opacity=1 ]   (221.95,77.5) -- (243.6,90) ;
\draw [shift={(235.89,85.55)}, rotate = 210] [color={rgb, 255:red, 208; green, 2; blue, 27 }  ,draw opacity=1 ][line width=0.75]    (6.56,-1.97) .. controls (4.17,-0.84) and (1.99,-0.18) .. (0,0) .. controls (1.99,0.18) and (4.17,0.84) .. (6.56,1.97)   ;

% Text Node
\draw (221.95,177.5) node  [font=\scriptsize] [align=left] {$\displaystyle \beta $};
% Text Node
\draw (198.3,177.5) node [anchor=east] [inner sep=0.75pt]  [font=\scriptsize] [align=left] {$\displaystyle 1$};
% Text Node
\draw (234.78,80.75) node [anchor=south west] [inner sep=0.75pt]  [font=\scriptsize] [align=left] {$\displaystyle 6$};
% Text Node
\draw (209.13,80.75) node [anchor=south east] [inner sep=0.75pt]  [font=\scriptsize] [align=left] {$\displaystyle 5$};
% Text Node
\draw (245.6,177.5) node [anchor=west] [inner sep=0.75pt]  [font=\scriptsize] [align=left] {$\displaystyle 4$};
% Text Node
\draw (234.78,155.75) node [anchor=south west] [inner sep=0.75pt]  [font=\scriptsize] [align=left] {$\displaystyle 3$};
% Text Node
\draw (209.13,155.75) node [anchor=south east] [inner sep=0.75pt]  [font=\scriptsize] [align=left] {$\displaystyle 2$};
% Text Node
\draw (221.95,74.5) node [anchor=south] [inner sep=0.75pt]  [font=\scriptsize] [align=left] {$\displaystyle \alpha $};

\end{tikzpicture}
    \caption{the ``smooth" and ``rough" boundary of quantum double model}
    \label{fig:boundary}
\end{figure}
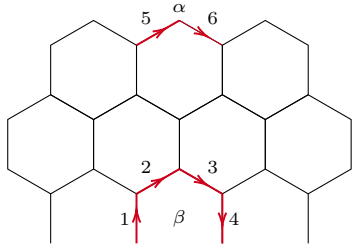

We first focus on two fundamental types of gapped boundaries, which generalize the smooth and rough boundaries of the Kitaev toric code~\cite{KITAEV20032} and play a central role in our construction~\cite{liGappedBoundariesKitaevs2025}. (In the language of the SymTFT framework, these two boundaries correspond exactly to the \(\Vecc_G\) and \(\cRep(G)\) boundaries introduced earlier.) Concretely, as shown in Fig.~\ref{fig:boundary}, the ``smooth'' boundary is characterized by a modified vertex term, while the ``rough'' boundary is characterized by a modified plaquette term. In our convention, the semi-vertex operators on the smooth boundary take the form
\(\hat{V}_\alpha=\delta_{g_5,g_6}\) (We will work straight forward on the projective of this projector.) and semi-plaquette operatores on the rough boundary take the form $\hat{P}_\beta(h)=\prod_{i=1}^4 L_h^{(i)}$, as shown in Fig.~\ref{fig:boundary}. 

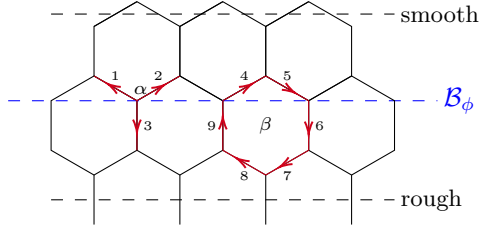
\begin{figure}
    \centering
    \begin{tikzpicture}[x=0.75pt,y=0.75pt,yscale=-1,xscale=1]

%Shape: Regular Polygon [id:dp6632657594749447] 
\draw   (128.39,138.33) -- (106.74,150.83) -- (85.09,138.33) -- (85.09,113.33) -- (106.74,100.83) -- (128.39,113.33) -- cycle ;
%Shape: Regular Polygon [id:dp10566388214482936] 
\draw   (171.69,138.33) -- (150.04,150.83) -- (128.39,138.33) -- (128.39,113.33) -- (150.04,100.83) -- (171.69,113.33) -- cycle ;
%Shape: Regular Polygon [id:dp41750001282657234] 
\draw   (193.34,175.83) -- (171.69,188.33) -- (150.04,175.83) -- (150.04,150.83) -- (171.69,138.33) -- (193.34,150.83) -- cycle ;
%Shape: Regular Polygon [id:dp9753661554809034] 
\draw   (150.04,175.83) -- (128.39,188.33) -- (106.74,175.83) -- (106.74,150.83) -- (128.39,138.33) -- (150.04,150.83) -- cycle ;
%Shape: Regular Polygon [id:dp6326888508964721] 
\draw   (106.74,175.83) -- (85.09,188.33) -- (63.44,175.83) -- (63.44,150.83) -- (85.09,138.33) -- (106.74,150.83) -- cycle ;
%Straight Lines [id:da12328154909120093] 
\draw [color={rgb, 255:red, 0; green, 0; blue, 255 }  ,draw opacity=1 ] [dash pattern={on 4.5pt off 4.5pt}]  (41.79,150.83) -- (258.29,150.83) ;
%Shape: Regular Polygon [id:dp2127501924535632] 
\draw   (236.64,175.83) -- (214.99,188.33) -- (193.34,175.83) -- (193.34,150.83) -- (214.99,138.33) -- (236.64,150.83) -- cycle ;
%Shape: Regular Polygon [id:dp07136269464879053] 
\draw   (214.99,138.33) -- (193.34,150.83) -- (171.69,138.33) -- (171.69,113.33) -- (193.34,100.83) -- (214.99,113.33) -- cycle ;
%Straight Lines [id:da5441806586556261] 
\draw [color={rgb, 255:red, 208; green, 2; blue, 27 }  ,draw opacity=1 ]   (106.74,150.83) -- (106.74,175.83) ;
\draw [shift={(106.74,166.93)}, rotate = 270] [color={rgb, 255:red, 208; green, 2; blue, 27 }  ,draw opacity=1 ][line width=0.75]    (6.56,-1.97) .. controls (4.17,-0.84) and (1.99,-0.18) .. (0,0) .. controls (1.99,0.18) and (4.17,0.84) .. (6.56,1.97)   ;
%Straight Lines [id:da874225891232131] 
\draw    (85.09,188.33) -- (85.09,213.33) ;
%Straight Lines [id:da4393139080574766] 
\draw    (128.39,188.33) -- (128.39,213.33) ;
%Straight Lines [id:da562706381212626] 
\draw    (171.69,188.33) -- (171.69,213.33) ;
%Straight Lines [id:da9263940293504959] 
\draw [color={rgb, 255:red, 208; green, 2; blue, 27 }  ,draw opacity=1 ]   (85.09,138.33) -- (106.74,150.83) ;
\draw [shift={(91.93,142.28)}, rotate = 30] [color={rgb, 255:red, 208; green, 2; blue, 27 }  ,draw opacity=1 ][line width=0.75]    (6.56,-1.97) .. controls (4.17,-0.84) and (1.99,-0.18) .. (0,0) .. controls (1.99,0.18) and (4.17,0.84) .. (6.56,1.97)   ;
%Straight Lines [id:da378931784101309] 
\draw [color={rgb, 255:red, 208; green, 2; blue, 27 }  ,draw opacity=1 ]   (106.74,150.83) -- (128.39,138.33) ;
\draw [shift={(120.68,142.78)}, rotate = 150] [color={rgb, 255:red, 208; green, 2; blue, 27 }  ,draw opacity=1 ][line width=0.75]    (6.56,-1.97) .. controls (4.17,-0.84) and (1.99,-0.18) .. (0,0) .. controls (1.99,0.18) and (4.17,0.84) .. (6.56,1.97)   ;
%Straight Lines [id:da5509293482513595] 
\draw [color={rgb, 255:red, 208; green, 2; blue, 27 }  ,draw opacity=1 ]   (150.04,150.83) -- (150.04,175.83) ;
\draw [shift={(150.04,158.73)}, rotate = 90] [color={rgb, 255:red, 208; green, 2; blue, 27 }  ,draw opacity=1 ][line width=0.75]    (6.56,-1.97) .. controls (4.17,-0.84) and (1.99,-0.18) .. (0,0) .. controls (1.99,0.18) and (4.17,0.84) .. (6.56,1.97)   ;
%Straight Lines [id:da2855529799642995] 
\draw [color={rgb, 255:red, 208; green, 2; blue, 27 }  ,draw opacity=1 ]   (150.04,175.83) -- (171.69,188.33) ;
\draw [shift={(156.88,179.78)}, rotate = 30] [color={rgb, 255:red, 208; green, 2; blue, 27 }  ,draw opacity=1 ][line width=0.75]    (6.56,-1.97) .. controls (4.17,-0.84) and (1.99,-0.18) .. (0,0) .. controls (1.99,0.18) and (4.17,0.84) .. (6.56,1.97)   ;
%Straight Lines [id:da347467006077943] 
\draw [color={rgb, 255:red, 208; green, 2; blue, 27 }  ,draw opacity=1 ]   (171.69,188.33) -- (193.34,175.83) ;
\draw [shift={(178.53,184.38)}, rotate = 330] [color={rgb, 255:red, 208; green, 2; blue, 27 }  ,draw opacity=1 ][line width=0.75]    (6.56,-1.97) .. controls (4.17,-0.84) and (1.99,-0.18) .. (0,0) .. controls (1.99,0.18) and (4.17,0.84) .. (6.56,1.97)   ;
%Straight Lines [id:da5575857183266836] 
\draw [color={rgb, 255:red, 208; green, 2; blue, 27 }  ,draw opacity=1 ]   (193.34,175.83) -- (193.34,150.83) ;
\draw [shift={(193.34,167.93)}, rotate = 270] [color={rgb, 255:red, 208; green, 2; blue, 27 }  ,draw opacity=1 ][line width=0.75]    (6.56,-1.97) .. controls (4.17,-0.84) and (1.99,-0.18) .. (0,0) .. controls (1.99,0.18) and (4.17,0.84) .. (6.56,1.97)   ;
%Straight Lines [id:da6493836739312008] 
\draw [color={rgb, 255:red, 208; green, 2; blue, 27 }  ,draw opacity=1 ]   (193.34,150.83) -- (171.69,138.33) ;
\draw [shift={(186.5,146.88)}, rotate = 210] [color={rgb, 255:red, 208; green, 2; blue, 27 }  ,draw opacity=1 ][line width=0.75]    (6.56,-1.97) .. controls (4.17,-0.84) and (1.99,-0.18) .. (0,0) .. controls (1.99,0.18) and (4.17,0.84) .. (6.56,1.97)   ;
%Straight Lines [id:da6337243935716672] 
\draw [color={rgb, 255:red, 208; green, 2; blue, 27 }  ,draw opacity=1 ]   (171.69,138.33) -- (150.04,150.83) ;
\draw [shift={(164.85,142.28)}, rotate = 150] [color={rgb, 255:red, 208; green, 2; blue, 27 }  ,draw opacity=1 ][line width=0.75]    (6.56,-1.97) .. controls (4.17,-0.84) and (1.99,-0.18) .. (0,0) .. controls (1.99,0.18) and (4.17,0.84) .. (6.56,1.97)   ;
%Straight Lines [id:da24493719375858736] 
\draw    (214.99,188.33) -- (214.99,213.33) ;
%Straight Lines [id:da6959665855352024] 
\draw    (171.69,113.33) -- (193.34,100.83) ;
%Straight Lines [id:da6376871313512747] 
\draw    (85.09,113.33) -- (106.74,100.83) ;
%Straight Lines [id:da4670514496886702] 
\draw  [dash pattern={on 4.5pt off 4.5pt}]  (63.44,107.08) -- (236.64,107.08) ;
%Straight Lines [id:da690389335905288] 
\draw    (150.04,100.83) -- (128.39,113.33) ;
%Straight Lines [id:da561251543358252] 
\draw  [dash pattern={on 4.5pt off 4.5pt}]  (63.44,200.83) -- (236.64,200.83) ;

% Text Node
\draw (260.29,150.83) node [anchor=west] [inner sep=0.75pt]  [font=\normalsize] [align=left] {$\displaystyle \textcolor[rgb]{0,0,1}{\mathcal{B}_{\phi }}$};
% Text Node
\draw (108.57,148.77) node [anchor=south] [inner sep=0.75pt]  [font=\scriptsize] [align=left] {$\displaystyle \alpha $};
% Text Node
\draw (171.69,163.33) node  [font=\scriptsize] [align=left] {$\displaystyle \beta $};
% Text Node
\draw (238.64,107.08) node [anchor=west] [inner sep=0.75pt]   [align=left] {smooth};
% Text Node
\draw (238.64,200.83) node [anchor=west] [inner sep=0.75pt]   [align=left] {rough};
% Text Node
\draw (95.91,141.58) node [anchor=south] [inner sep=0.75pt]  [font=\tiny] [align=left] {$\displaystyle 1$};
% Text Node
\draw (148.04,163.33) node [anchor=east] [inner sep=0.75pt]  [font=\tiny] [align=left] {$\displaystyle 9$};
% Text Node
\draw (160.87,185.08) node [anchor=north] [inner sep=0.75pt]  [font=\tiny] [align=left] {$\displaystyle 8$};
% Text Node
\draw (182.52,185.08) node [anchor=north] [inner sep=0.75pt]  [font=\tiny] [align=left] {$\displaystyle 7$};
% Text Node
\draw (182.52,141.58) node [anchor=south] [inner sep=0.75pt]  [font=\tiny] [align=left] {$\displaystyle 5$};
% Text Node
\draw (160.87,141.58) node [anchor=south] [inner sep=0.75pt]  [font=\tiny] [align=left] {$\displaystyle 4$};
% Text Node
\draw (108.74,163.33) node [anchor=west] [inner sep=0.75pt]  [font=\tiny] [align=left] {$\displaystyle 3$};
% Text Node
\draw (117.56,141.58) node [anchor=south] [inner sep=0.75pt]  [font=\tiny] [align=left] {$\displaystyle 2$};
% Text Node
\draw (195.34,163.33) node [anchor=west] [inner sep=0.75pt]  [font=\tiny] [align=left] {$\displaystyle 6$};

\end{tikzpicture}
    \caption{the domain wall $\mathcal{B}_\phi$ between smooth and rough boundary}
    \label{fig:domain wall}
\end{figure}

Another key defect we will use is a domain wall labeled by an endomorphism \(\phi\in\Endo(G)\), denoted by \(\mathcal{B}_\phi\). Crossing \(\mathcal{B}_\phi\), the local constraints of the quantum double model are \(\phi\)-twisted: vertex and plaquette terms adjacent to the wall are modified so that the group multiplication rules and gauge transformations are acted on by \(\phi\) on one side of the wall.

For instance, in the geometry of Fig.~\ref{fig:domain wall}, the vertex constraint at \(\alpha\) is replaced by
\[
\hat{V}_\alpha=\delta_{e,\phi(g_1g_2)g_3},
\]
and a plaquette term crossing the wall acts as
\[
\hat{P}_\beta(h)=L_h^{(4)}L_h^{(5)}L_{\phi(h)}^{(6)}L_{\phi(h)}^{(7)}L_{\phi(h)}^{(8)}L_{\phi(h)}^{(9)}.
\]
Intuitively, edges on one side of the wall transform by \(h\), while edges on the other side transform by \(\phi(h)\).

\begin{figure}[h]
    \centering
    \begin{tikzpicture}[x=0.75pt,y=0.75pt,yscale=-1,xscale=1]

%Straight Lines [id:da2650289686854458] 
\draw [color={rgb, 255:red, 0; green, 0; blue, 255 }  ,draw opacity=1 ] [dash pattern={on 4.5pt off 4.5pt}]  (60.05,121.5) -- (319.86,121.5) ;
%Straight Lines [id:da28879463670059513] 
\draw    (125,109) -- (146.65,121.5) ;
%Straight Lines [id:da9538654751120809] 
\draw    (146.65,121.5) -- (146.65,146.5) ;
%Straight Lines [id:da5232086443277826] 
\draw    (146.65,121.5) -- (168.3,109) ;

%Straight Lines [id:da6549226902882004] 
\draw    (81.7,109) -- (103.35,121.5) ;
%Straight Lines [id:da17073578970428704] 
\draw    (103.35,121.5) -- (103.35,146.5) ;
%Straight Lines [id:da8173370794994735] 
\draw    (103.35,121.5) -- (125,109) ;

%Straight Lines [id:da08630282056671335] 
\draw    (211.6,109) -- (233.25,121.5) ;
%Straight Lines [id:da7919315093625023] 
\draw    (233.25,121.5) -- (233.25,146.5) ;
%Straight Lines [id:da8133567630206681] 
\draw    (233.25,121.5) -- (254.9,109) ;

%Straight Lines [id:da43057141279563826] 
\draw    (168.3,109) -- (189.95,121.5) ;
%Straight Lines [id:da00661064580432924] 
\draw    (189.95,121.5) -- (189.95,146.5) ;
%Straight Lines [id:da12581995862520734] 
\draw    (189.95,121.5) -- (211.6,109) ;

%Straight Lines [id:da8690306796069706] 
\draw    (254.9,109) -- (276.55,121.5) ;
%Straight Lines [id:da6798761144448926] 
\draw    (276.55,121.5) -- (276.55,146.5) ;
%Straight Lines [id:da6296263715987513] 
\draw    (276.55,121.5) -- (298.21,109) ;

%Straight Lines [id:da06291798528987547] 
\draw [color={rgb, 255:red, 208; green, 2; blue, 27 }  ,draw opacity=1 ]   (168.3,109) -- (189.95,121.5) ;
\draw [shift={(182.24,117.05)}, rotate = 210] [color={rgb, 255:red, 208; green, 2; blue, 27 }  ,draw opacity=1 ][line width=0.75]    (6.56,-1.97) .. controls (4.17,-0.84) and (1.99,-0.18) .. (0,0) .. controls (1.99,0.18) and (4.17,0.84) .. (6.56,1.97)   ;
%Straight Lines [id:da7080402653717153] 
\draw [color={rgb, 255:red, 208; green, 2; blue, 27 }  ,draw opacity=1 ]   (189.95,121.5) -- (211.6,109) ;
\draw [shift={(203.89,113.45)}, rotate = 150] [color={rgb, 255:red, 208; green, 2; blue, 27 }  ,draw opacity=1 ][line width=0.75]    (6.56,-1.97) .. controls (4.17,-0.84) and (1.99,-0.18) .. (0,0) .. controls (1.99,0.18) and (4.17,0.84) .. (6.56,1.97)   ;
%Straight Lines [id:da1886339081354853] 
\draw [color={rgb, 255:red, 208; green, 2; blue, 27 }  ,draw opacity=1 ]   (189.95,121.5) -- (189.95,146.5) ;
\draw [shift={(189.95,137.6)}, rotate = 270] [color={rgb, 255:red, 208; green, 2; blue, 27 }  ,draw opacity=1 ][line width=0.75]    (6.56,-1.97) .. controls (4.17,-0.84) and (1.99,-0.18) .. (0,0) .. controls (1.99,0.18) and (4.17,0.84) .. (6.56,1.97)   ;
%Straight Lines [id:da2960957984432627] 
\draw [color={rgb, 255:red, 208; green, 2; blue, 27 }  ,draw opacity=1 ]   (146.65,121.5) -- (168.3,109) ;
\draw [shift={(160.59,113.45)}, rotate = 150] [color={rgb, 255:red, 208; green, 2; blue, 27 }  ,draw opacity=1 ][line width=0.75]    (6.56,-1.97) .. controls (4.17,-0.84) and (1.99,-0.18) .. (0,0) .. controls (1.99,0.18) and (4.17,0.84) .. (6.56,1.97)   ;
%Straight Lines [id:da9037578912121101] 
\draw [color={rgb, 255:red, 208; green, 2; blue, 27 }  ,draw opacity=1 ]   (211.6,109) -- (233.25,121.5) ;
\draw [shift={(225.55,117.05)}, rotate = 210] [color={rgb, 255:red, 208; green, 2; blue, 27 }  ,draw opacity=1 ][line width=0.75]    (6.56,-1.97) .. controls (4.17,-0.84) and (1.99,-0.18) .. (0,0) .. controls (1.99,0.18) and (4.17,0.84) .. (6.56,1.97)   ;
%Straight Lines [id:da0011706828982485096] 
\draw [color={rgb, 255:red, 208; green, 2; blue, 27 }  ,draw opacity=1 ]   (146.65,146.5) -- (146.65,121.5) ;
\draw [shift={(146.65,138.6)}, rotate = 270] [color={rgb, 255:red, 208; green, 2; blue, 27 }  ,draw opacity=1 ][line width=0.75]    (6.56,-1.97) .. controls (4.17,-0.84) and (1.99,-0.18) .. (0,0) .. controls (1.99,0.18) and (4.17,0.84) .. (6.56,1.97)   ;

% Text Node
\draw (168.3,106) node [anchor=south] [inner sep=0.75pt]  [font=\scriptsize] [align=left] {$\displaystyle g_{i}$};
% Text Node
\draw (211.6,106) node [anchor=south] [inner sep=0.75pt]  [font=\scriptsize] [align=left] {$\displaystyle g_{i+1}$};
% Text Node
\draw (191.95,134) node [anchor=west] [inner sep=0.75pt]  [font=\scriptsize] [align=left] {$\displaystyle g_{i+\tfrac{1}{2}}$};
% Text Node
\draw (144.65,134) node [anchor=east] [inner sep=0.75pt]  [font=\scriptsize] [align=left] {$\displaystyle g_{i-\tfrac{1}{2}}$};
% Text Node
\draw (321.86,121.5) node [anchor=west] [inner sep=0.75pt]  [font=\footnotesize,color={rgb, 255:red, 0; green, 0; blue, 255 }  ,opacity=1 ] [align=left] {$\displaystyle \mathcal{B}_{\phi }$};
% Text Node
\draw (74.7,125.5) node [anchor=north west][inner sep=0.75pt]   [align=left] {...};
% Text Node
\draw (291.21,125.5) node [anchor=north west][inner sep=0.75pt]   [align=left] {...};
\end{tikzpicture}
    \caption{the contracted $1+1d$ model}
    \label{fig:1d contract}
\end{figure}
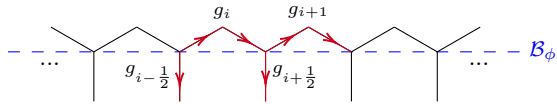

Combining the two gapped boundaries discussed above with the \(\phi\)-domain wall, we now take the quasi-1D limit by shrinking the width of the configuration in Fig.~\ref{fig:domain wall} to a single strip (one ``grid''), yielding an effective \((1+1)\)-dimensional model shown in Fig.~\ref{fig:1d contract}. In this limit the Hamiltonian is generated by two families of mutually commuting operators, which we interpret as the effective vertex and plaquette terms along the chain:
\[
    \hat{V}_i \coloneq \delta_{e,\phi(\bar{g}_i)\,g_{i+\frac{1}{2}}\,\phi(g_{i+1})}, \quad
    \hat{P}_i \coloneq \frac{1}{|G|}\sum_{h\in G}R_{\phi(\bar{h})}^{\left(i-\frac{1}{2}\right)}\,L_h^{(i)}\,L_{\phi(h)}^{\left(i+\frac{1}{2}\right)}.
\]
Here the half-integer labels refer to edge degrees of freedom between neighboring plaquette, and \(L\) and \(R\) denote the left and right actions introduced above. We have also identified the two adjacent edges meeting at each semi-vertex, so that each integer position carries a single effective edge degree of freedom.

The contraction form a $2+1d$ theory to a $1+1d$ one seems redundancy. However, this interpretation encodes the ideology of SymTFT which is essential for the classification and realization of SPT phases.

We now introduce two families of operators---an \(X\)-type and a \(Z\)-type---which generalize the Pauli operators appearing in the group-based cluster model~\cite{PhysRevX.15.011058}. They are defined by
\begin{equation*}
    \overset{\leftarrow}{X}_h=\sum_{g\in G}|g\bar{h}\rangle\langle g|,\quad \overset{\rightarrow}{X}_h=\sum_{g\in G} |hg\rangle\langle g|, \quad Z_\Gamma=\sum_{g\in G}\Gamma(g)|g\rangle\langle g|,
\end{equation*}
for any \(h\in G\) and any representation \(\Gamma\) of \(G\). Here \(\overset{\rightarrow}{X}_h\) and \(\overset{\leftarrow}{X}_h\) implement the left and right regular actions (i.e., group multiplication on the ket from the left or from the right), while \(Z_\Gamma\) is diagonal in the group-element basis and weights each basis state \(|g\rangle\) by the representation matrix \(\Gamma(g)\).

In terms of these operators, the quasi-1D vertex and plaquette constraints derived above can be rewritten as
\begin{align*}
    \hat{V_i}=\frac{1}{|G|}\sum_{\Gamma\in\irrep(G)}d_\Gamma\Tr\left(Z_{\phi^*\Gamma}^{\dagger(i)}\cdot Z_\Gamma^{(i+\frac{1}{2})}\cdot Z_{\phi^*\Gamma}^{(i+1)}\right),\quad
    \hat{P_i}=\frac{1}{|G|}\sum_{h\in G}\overset{\leftarrow}{X}_{\phi(h)}^{(i-\frac{1}{2})} \overset{\rightarrow}{X}_h^{(i)}\overset{\rightarrow}{X}_{\phi(h)}^{(i+\frac{1}{2})}.
\end{align*}
The first expression is a character/Fourier resolution of the group-valued constraint appearing in \(\hat{V_i}\): the sum over \(\Gamma\in\irrep(G)\), weighted by \(d_\Gamma\), enforces the corresponding \(\delta\)-type condition through the standard orthogonality relations of irreducible representations. The second expression makes the ``twisted'' gauge transformation structure manifest: the degree of freedom at the integer site \(i\) transforms by \(h\), while the neighboring half-integer degrees of freedom transform by \(\phi(h)\), with the left/right arrows recording whether the multiplication acts on the ket from the left or from the right.

Therefore, the Hamiltonian of the effective \((1+1)\)-dimensional model reads
\begin{equation}
    H=-\frac{1}{|G|}\sum_i\left[\sum_{\Gamma\in \irrep(G)}d_\Gamma\Tr\left(Z_{\phi^*\Gamma}^{\dagger(i)}\cdot Z_\Gamma^{(i+\frac{1}{2})}\cdot Z_{\phi^*\Gamma}^{(i+1)}\right)+\sum_{h\in G}\overset{\leftarrow}{X}_{\phi(h)}^{(i-\frac{1}{2})} \overset{\rightarrow}{X}_h^{(i)}\overset{\rightarrow}{X}_{\phi(h)}^{(i+\frac{1}{2})}\right].
\end{equation}
By construction, all terms in this Hamiltonian commute with one another, so the model is exactly solvable and its ground states can be obtained by imposing the simultaneous \(+1\) eigenvalue conditions of all \(\hat{V_i}\) and \(\hat{P_i}\). In particular, for an $N$ sites open chain, one can write an explicit ground state in the group-element basis as
\begin{equation}
    |\Omega\rangle=\sum_{g_2,\ldots,g_{N-1}}|g_1\rangle \otimes|\phi(g_1\bar{g_2})\rangle \otimes |g_2\rangle \otimes |\phi(g_2\bar{g_3})\rangle \otimes ...\otimes|g_N\rangle.
\end{equation}

This expression makes the structure of the quasi-1D constraints transparent: the integer sites carry the variables \(g_i\), while the half-integer sites store the \(\phi\)-twisted ``differences'' between neighboring group elements. Note that the two terminal $g_1$ and $g_N$ are arbitrary, thus the ground space is $|G|^2$ degeneracy.

Two limits are particularly instructive. When \(\phi = e\), the half-integer degrees of freedom are pinned to the identity element, and the ground state reduces to
\[
|\Omega\rangle=\sum_{g_2,\ldots,g_{N-1}}|g_1\rangle \otimes|e\rangle \otimes |g_2\rangle \otimes |e\rangle \otimes ...\otimes|g_N\rangle,
\]
which is a simple product state. In contrast, when \(\phi = \mathrm{id}\), the half-integer sites record the untwisted relative group elements \(g_i\bar{g}_{i+1}\), and we obtain
\[
|\Omega\rangle=\sum_{g_2,\ldots,g_{N-1}}|g_1\rangle \otimes|g_1\bar{g_2}\rangle \otimes |g_2\rangle \otimes |g_2\bar{g_3}\rangle \otimes ...\otimes|g_N\rangle,
\]
which is precisely the group-based cluster state introduced in~\cite{PhysRevX.15.011058}.

The global symmetry act on the chain is characterized by two families of ribbon operators corresponding to the $G$ and $\Rep(G)$ symmetries respectively, which are:
\begin{equation}
\begin{aligned}
    A_g=&\vcenter{\hbox{
    \begin{tikzpicture}[x=0.75pt,y=0.75pt,yscale=-1,xscale=1]

%Straight Lines [id:da9367906268722342] 
\draw [color={rgb, 255:red, 0; green, 0; blue, 255 }  ,draw opacity=1 ] [dash pattern={on 4.5pt off 4.5pt}]  (60.05,121.5) -- (319.86,121.5) ;
%Straight Lines [id:da2177425320075551] 
\draw    (125,109) -- (146.65,121.5) ;
%Straight Lines [id:da972977001309482] 
\draw    (146.65,121.5) -- (146.65,146.5) ;
%Straight Lines [id:da16244517604100772] 
\draw    (146.65,121.5) -- (168.3,109) ;

%Straight Lines [id:da481519892372268] 
\draw    (81.7,109) -- (103.35,121.5) ;
%Straight Lines [id:da4892958159802401] 
\draw    (103.35,121.5) -- (103.35,146.5) ;
%Straight Lines [id:da2421595137433865] 
\draw    (103.35,121.5) -- (125,109) ;

%Straight Lines [id:da9066936724570167] 
\draw    (211.6,109) -- (233.25,121.5) ;
%Straight Lines [id:da42494733372528126] 
\draw    (233.25,121.5) -- (233.25,146.5) ;
%Straight Lines [id:da8228043720433112] 
\draw    (233.25,121.5) -- (254.9,109) ;

%Straight Lines [id:da179065147855742] 
\draw    (168.3,109) -- (189.95,121.5) ;
%Straight Lines [id:da4872153867083744] 
\draw    (189.95,121.5) -- (189.95,146.5) ;
%Straight Lines [id:da17104359267553337] 
\draw    (189.95,121.5) -- (211.6,109) ;

%Straight Lines [id:da450191289930686] 
\draw    (254.9,109) -- (276.55,121.5) ;
%Straight Lines [id:da5680634953243303] 
\draw    (276.55,121.5) -- (276.55,146.5) ;
%Straight Lines [id:da10755824756729404] 
\draw    (276.55,121.5) -- (298.21,109) ;

%Straight Lines [id:da8454859513842926] 
\draw [color={rgb, 255:red, 208; green, 2; blue, 27 }  ,draw opacity=1 ][line width=1.5]    (103.35,121.5) -- (81.7,109) ;
\draw [shift={(87.85,112.55)}, rotate = 30] [color={rgb, 255:red, 208; green, 2; blue, 27 }  ,draw opacity=1 ][line width=1.5]    (8.53,-2.57) .. controls (5.42,-1.09) and (2.58,-0.23) .. (0,0) .. controls (2.58,0.23) and (5.42,1.09) .. (8.53,2.57)   ;
%Straight Lines [id:da11953646083543845] 
\draw [color={rgb, 255:red, 208; green, 2; blue, 27 }  ,draw opacity=1 ][line width=1.5]    (125,109) -- (103.35,121.5) ;
\draw [shift={(109.5,117.95)}, rotate = 330] [color={rgb, 255:red, 208; green, 2; blue, 27 }  ,draw opacity=1 ][line width=1.5]    (8.53,-2.57) .. controls (5.42,-1.09) and (2.58,-0.23) .. (0,0) .. controls (2.58,0.23) and (5.42,1.09) .. (8.53,2.57)   ;
%Straight Lines [id:da6845487582109798] 
\draw    (146.65,121.5) -- (125,109) ;
%Straight Lines [id:da8151218595706177] 
\draw [color={rgb, 255:red, 208; green, 2; blue, 27 }  ,draw opacity=1 ][line width=1.5]    (146.65,121.5) -- (125,109) ;
\draw [shift={(131.15,112.55)}, rotate = 30] [color={rgb, 255:red, 208; green, 2; blue, 27 }  ,draw opacity=1 ][line width=1.5]    (8.53,-2.57) .. controls (5.42,-1.09) and (2.58,-0.23) .. (0,0) .. controls (2.58,0.23) and (5.42,1.09) .. (8.53,2.57)   ;
%Straight Lines [id:da5196959505259771] 
\draw [color={rgb, 255:red, 208; green, 2; blue, 27 }  ,draw opacity=1 ][line width=1.5]    (189.95,121.5) -- (168.3,109) ;
\draw [shift={(174.45,112.55)}, rotate = 30] [color={rgb, 255:red, 208; green, 2; blue, 27 }  ,draw opacity=1 ][line width=1.5]    (8.53,-2.57) .. controls (5.42,-1.09) and (2.58,-0.23) .. (0,0) .. controls (2.58,0.23) and (5.42,1.09) .. (8.53,2.57)   ;
%Straight Lines [id:da1847841058081059] 
\draw [color={rgb, 255:red, 208; green, 2; blue, 27 }  ,draw opacity=1 ][line width=1.5]    (233.25,121.5) -- (211.6,109) ;
\draw [shift={(217.75,112.55)}, rotate = 30] [color={rgb, 255:red, 208; green, 2; blue, 27 }  ,draw opacity=1 ][line width=1.5]    (8.53,-2.57) .. controls (5.42,-1.09) and (2.58,-0.23) .. (0,0) .. controls (2.58,0.23) and (5.42,1.09) .. (8.53,2.57)   ;
%Straight Lines [id:da16588943979389892] 
\draw [color={rgb, 255:red, 208; green, 2; blue, 27 }  ,draw opacity=1 ][line width=1.5]    (276.55,121.5) -- (254.9,109) ;
\draw [shift={(261.05,112.55)}, rotate = 30] [color={rgb, 255:red, 208; green, 2; blue, 27 }  ,draw opacity=1 ][line width=1.5]    (8.53,-2.57) .. controls (5.42,-1.09) and (2.58,-0.23) .. (0,0) .. controls (2.58,0.23) and (5.42,1.09) .. (8.53,2.57)   ;
%Straight Lines [id:da4664391722184188] 
\draw [color={rgb, 255:red, 208; green, 2; blue, 27 }  ,draw opacity=1 ][line width=1.5]    (168.3,109) -- (146.65,121.5) ;
\draw [shift={(152.8,117.95)}, rotate = 330] [color={rgb, 255:red, 208; green, 2; blue, 27 }  ,draw opacity=1 ][line width=1.5]    (8.53,-2.57) .. controls (5.42,-1.09) and (2.58,-0.23) .. (0,0) .. controls (2.58,0.23) and (5.42,1.09) .. (8.53,2.57)   ;
%Straight Lines [id:da48788597400442923] 
\draw [color={rgb, 255:red, 208; green, 2; blue, 27 }  ,draw opacity=1 ][line width=1.5]    (211.6,109) -- (189.95,121.5) ;
\draw [shift={(196.1,117.95)}, rotate = 330] [color={rgb, 255:red, 208; green, 2; blue, 27 }  ,draw opacity=1 ][line width=1.5]    (8.53,-2.57) .. controls (5.42,-1.09) and (2.58,-0.23) .. (0,0) .. controls (2.58,0.23) and (5.42,1.09) .. (8.53,2.57)   ;
%Straight Lines [id:da8599736037227078] 
\draw [color={rgb, 255:red, 208; green, 2; blue, 27 }  ,draw opacity=1 ][line width=1.5]    (254.9,109) -- (233.25,121.5) ;
\draw [shift={(239.4,117.95)}, rotate = 330] [color={rgb, 255:red, 208; green, 2; blue, 27 }  ,draw opacity=1 ][line width=1.5]    (8.53,-2.57) .. controls (5.42,-1.09) and (2.58,-0.23) .. (0,0) .. controls (2.58,0.23) and (5.42,1.09) .. (8.53,2.57)   ;
%Straight Lines [id:da5737238197565928] 
\draw [color={rgb, 255:red, 208; green, 2; blue, 27 }  ,draw opacity=1 ][line width=1.5]    (298.21,109) -- (276.55,121.5) ;
\draw [shift={(282.7,117.95)}, rotate = 330] [color={rgb, 255:red, 208; green, 2; blue, 27 }  ,draw opacity=1 ][line width=1.5]    (8.53,-2.57) .. controls (5.42,-1.09) and (2.58,-0.23) .. (0,0) .. controls (2.58,0.23) and (5.42,1.09) .. (8.53,2.57)   ;

% Text Node
\draw (321.86,121.5) node [anchor=west] [inner sep=0.75pt]  [font=\footnotesize,color={rgb, 255:red, 0; green, 0; blue, 255 }  ,opacity=1 ] [align=left] {$\displaystyle \mathcal{B}_{\phi }$};
% Text Node
\draw (74.7,125.5) node [anchor=north west][inner sep=0.75pt]   [align=left] {...};
% Text Node
\draw (291.21,125.5) node [anchor=north west][inner sep=0.75pt]   [align=left] {...};
% Text Node
\draw (198.78,112.25) node [anchor=south east] [inner sep=0.75pt]  [font=\small] [align=left] {$\displaystyle g$};

\end{tikzpicture}
    }} = \prod_i  \overset{\leftarrow}{X}_g^{(i)},\\
    B_{\Gamma_{\alpha\beta}}=& \vcenter{\hbox{
    \begin{tikzpicture}[x=0.75pt,y=0.75pt,yscale=-1,xscale=1]

%Straight Lines [id:da21831718709670334] 
\draw [color={rgb, 255:red, 0; green, 0; blue, 255 }  ,draw opacity=1 ] [dash pattern={on 4.5pt off 4.5pt}]  (60.05,121.5) -- (319.86,121.5) ;
%Straight Lines [id:da4177456308694165] 
\draw    (125,109) -- (146.65,121.5) ;
%Straight Lines [id:da6506172149630641] 
\draw    (146.65,121.5) -- (146.65,146.5) ;
%Straight Lines [id:da6247117908547151] 
\draw    (146.65,121.5) -- (168.3,109) ;

%Straight Lines [id:da6615644005681827] 
\draw    (81.7,109) -- (103.35,121.5) ;
%Straight Lines [id:da34181089896512606] 
\draw    (103.35,121.5) -- (103.35,146.5) ;
%Straight Lines [id:da06557606816439832] 
\draw    (103.35,121.5) -- (125,109) ;

%Straight Lines [id:da07347583085333065] 
\draw    (211.6,109) -- (233.25,121.5) ;
%Straight Lines [id:da38807326687411103] 
\draw    (233.25,121.5) -- (233.25,146.5) ;
%Straight Lines [id:da30930259764459067] 
\draw    (233.25,121.5) -- (254.9,109) ;

%Straight Lines [id:da6884199039651018] 
\draw    (168.3,109) -- (189.95,121.5) ;
%Straight Lines [id:da43489475117940246] 
\draw    (189.95,121.5) -- (189.95,146.5) ;
%Straight Lines [id:da6075247847181472] 
\draw    (189.95,121.5) -- (211.6,109) ;

%Straight Lines [id:da5102769260970587] 
\draw    (254.9,109) -- (276.55,121.5) ;
%Straight Lines [id:da32266888936668714] 
\draw    (276.55,121.5) -- (276.55,146.5) ;
%Straight Lines [id:da38614579424385187] 
\draw    (276.55,121.5) -- (298.21,109) ;

%Straight Lines [id:da7941862743481364] 
\draw    (146.65,121.5) -- (125,109) ;
%Straight Lines [id:da4213648028385476] 
\draw [color={rgb, 255:red, 65; green, 117; blue, 5 }  ,draw opacity=1 ][line width=1.5]    (81.7,134) -- (298.21,134) ;
\draw [shift={(183.55,134)}, rotate = 0] [color={rgb, 255:red, 65; green, 117; blue, 5 }  ,draw opacity=1 ][line width=1.5]    (8.53,-2.57) .. controls (5.42,-1.09) and (2.58,-0.23) .. (0,0) .. controls (2.58,0.23) and (5.42,1.09) .. (8.53,2.57)   ;

% Text Node
\draw (321.86,121.5) node [anchor=west] [inner sep=0.75pt]  [font=\footnotesize,color={rgb, 255:red, 0; green, 0; blue, 255 }  ,opacity=1 ] [align=left] {$\displaystyle \mathcal{B}_{\phi }$};
% Text Node
\draw (74.7,125.5) node [anchor=north west][inner sep=0.75pt]   [align=left] {...};
% Text Node
\draw (291.21,125.5) node [anchor=north west][inner sep=0.75pt]   [align=left] {...};
% Text Node
\draw (187.95,137) node [anchor=north east] [inner sep=0.75pt]  [font=\small] [align=left] {$\displaystyle \Gamma $};

\end{tikzpicture}
    }}=\left(\prod_i Z_\Gamma^{(i+\frac{1}{2})}\right)_{\alpha\beta}.
\end{aligned}
\end{equation}

Within the ground-state subspace, one can directly check that the global symmetry acts effectively as operators supported only at the two ends of the chain. In particular, the symmetry generators reduce to the boundary actions
\begin{equation}
    A_g=\overset{\leftarrow}{X}_g^{(1)}\overset{\leftarrow}{X}_g^{(N)},\quad B_{\Gamma_{\alpha\beta}}=\left(Z_{\phi^*\Gamma}^{(1)}\cdot Z_{\phi^*\Gamma}^{\dagger(N)}\right)_{\alpha \beta}.
\end{equation}
That is, although the symmetry is defined microscopically as a global transformation, its action on the low-energy (ground) subspace is represented entirely by local operators at the left and right boundaries. This ``symmetry fractionalization'' onto the ends can be viewed as the appearance of protected edge modes, which is an essential diagnostic of SPT order.

Notice that the edge modes of $G$ symmetry and $\Rep(G)$ symmetry commute within each family but do not commute with one another, For example,
\begin{equation}
    \overset{\leftarrow}{X}_g^{(1)} Z_{\phi^*\Gamma}^{(1)}=\left[Z_{\phi^*\Gamma}^{(1)}\cdot\Gamma(\phi g)\right] \overset{\leftarrow}{X}_g^{(1)}.
\end{equation}
However, the total edge operators – the product of the left and right operators – still commute. This precise serves as an interpretation of what is called ``intrinsically mixed".

Finally, we emphasize that the discussion above assumes that both endpoints are integer sites (i.e., horizontal edges in our quasi-1D geometry). The case in which the endpoints lie on half-integer sites (vertical edges) is completely analogous, with the corresponding boundary operators obtained by the same reduction procedure.

\section{Classify of $1+1$d $\Rep(G) \times G$ SPT}
Having constructed, in Sec.\ref{Construction of the models}, a family of quasi-1D Hamiltonian from a modified quantum double model and identified its ground state as a \(\phi\)-twisted group-based cluster state, we now turn to a purely categorical classification of the corresponding \((1+1)\)D SPT phases. The goal of this section is to explain why the lattice constructions obtained from different domain-wall twists \(\phi\in\Endo(G)\) exhaust all \(\Rep(G)\times G\) (equivalently \(\cRep(G)\boxtimes\Vecc_G\)) SPT phases in \((1+1)\) dimensions, and to provide a compact invariant that distinguishes them.

The guiding principle is that \((1+1)\)D bosonic SPT phases with categorical symmetry \(\mathcal{C}\) are classified by \(\mathcal{C}\)-module categories over \(\Vecc\), or equivalently by fiber functors \(\mathcal{C}\to\Vecc\)~\cite{thorngrenFusionCategorySymmetry2024}. In the present problem the symmetry category is
\[
\mathcal{H}\coloneqq \cRep(G)\boxtimes \Vecc_G,
\]
so we seek all \(\mathcal{H}\)-module structures on \(\Vecc\) and identify the associated monoidal functors. Physically, this classification organizes the possible ways in which the \(\Rep(G)\) and \(G\) symmetry defects can be consistently ``glued'' on the boundary; mathematically, it characterizes the admissible condensations in the stacked bulk \(\Zc(\mathcal{H})\simeq \mathcal{D}_G^2\) that correspond to SPT (i.e., short-range entangled) phases.

This section is structured as follows. In Sec.~III~A we classify \(\cRep(G)\boxtimes\Vecc_G\)-module structures on \(\Vecc\) by translating the problem into one about bimodules over suitable algebra objects inside \(\Vecc_{G\times G}\). In Sec.~III~B we compute the corresponding associator data and extract the fiber functor explicitly, showing that the resulting invariant depends only on the class of \(\phi\) modulo inner automorphisms. These results will then be matched, in Sec.~IV, with the domain-wall picture and the condensable algebra \(\mathcal{A}_\phi\) in \(\mathcal{D}_G^2\), thereby connecting the categorical classification to the lattice realization constructed earlier.

\subsection{$\Rep(G) \times G$-module structures on $\Vecc$}\label{Rep(G) times G-module structures on Vecc}

Constructing the desired module structure directly from the axiomatic definition of an \(\mathcal{H}\)-module category is cumbersome. Instead, we use a standard identification that realizes \(\mathcal{H}\) as a bimodule category inside \(\mathcal{G}\). Let
\[
G_1 \coloneqq \{(g,e)\mid g\in G\}\subseteq G\times G,
\]
and denote by \(\hat{G_1}\) the corresponding group algebra, viewed as an algebra object in \(\mathcal{G}\). Then \(\mathcal{H}\) is equivalent to the category of \(\hat{G_1}\)-bimodules in \(\mathcal{G}\). More precisely, there is an equivalence
\begin{align*}
    \mathcal{H} \xrightarrow{\sim}& \bimod_{\mathcal{G}}(\hat{G_1});\\
    \Gamma\times g \mapsto& \left(\bigoplus_h (h,g)^{\oplus \dim(\Gamma)},p,q\right) \equiv {}_g\Gamma,
\end{align*}
where \(p\) and \(q\) denote the left and right \(\hat{G_1}\)-module structures, respectively.

Concretely, fixing a basis \(\{\hat{e}(h,g)_i\}\) of \({}_g\Gamma\) (with \(i=1,\dots,d_\Gamma\), where \(d_\Gamma\) is the dimension of \(\Gamma\)), the left action \(p\) is defined by
\begin{align*}
    \hat{G_1} \otimes {}_g\Gamma \to& {}_g\Gamma;\\
    (h,e) \otimes \hat{e}(h_0,g_0)_i \mapsto & \sum_j\Gamma^{-1}(h)_{ij}\hat{e}(hh_0,g_0)_j,
\end{align*}
while the right action \(q\) is given by
\begin{align*}
    {}_g\Gamma \otimes \hat{G_1} \to& {}_g\Gamma;\\
    \hat{e}(h_0,g_0)_i \otimes (h,e) \mapsto& \hat{e}(h_0h,g_0)_i.
\end{align*}
In other words, \(\hat{G_1}\) acts on the left by left multiplication on the first group component, accompanied by the representation matrix \(\Gamma^{-1}(h)\) on the multiplicity space, whereas the right action is the regular right multiplication on the same component and is trivial on the multiplicity index. 

In diagram:
\begin{equation}
    \vcenter{\hbox{
    \begin{tikzpicture}
        \insertion{(-0.6,-0.8)}{};
        \insertion{(0,-0.8)}{i};
        \insertion{(0.6,-0.8)}{};
        \projection{(0,0.8)}{j};
        \LeftModuleAction{(0,0.4)};
        \RightModuleAction{(0,0)};
        \draw[thick, decoration = {markings, mark=at position 0.2 with {\arrow[scale=1]{stealth}; \node[left, font=\scriptsize]{${}_g\Gamma$};}}, postaction=decorate] (0,-0.8) -- (0,0.8);
        \draw[algebra, thick, decoration = {markings, mark=at position 0.6 with {\arrow[scale=1]{stealth}; \node[left, font=\scriptsize]{$\hat{G_1}$};}}, postaction=decorate] (-0.6,-0.8) to[in=240, out=90]  (-0.12,0.4);
        \draw[algebra, thick, decoration = {markings, mark=at position 0.6 with {\arrow[scale=1]{stealth}; \node[right, font=\scriptsize]{$\hat{G_1}$};}}, postaction=decorate] (0.6,-0.8)   to[in=300, out=90] (0.12,0);
        \draw[simple] (-0.6,-1.2) -- (-0.6,-0.8);
        \draw[simple] (0,-1.2) -- (0,-0.8);
        \draw[simple] (0.6,-1.2) -- (0.6,-0.8);
        \draw[simple] (0,0.8) -- (0,1.2);
        \draw[](-0.8,-1.1) node[below]{\scriptsize$(g_2,1)$};
        \draw[](0,-1.1) node[below] {\scriptsize$(g,h)$};
        \draw[](0.8,-1.1) node[below] {\scriptsize$(g_1,1)$};
        \draw[](0,1.1) node[above] {\scriptsize$(g_2gg_1,h)$};
    \end{tikzpicture}
    }}
    =
    \vcenter{\hbox{
    \begin{tikzpicture}
        \insertion{(-0.6,-0.8)}{};
        \insertion{(0,-0.8)}{i};
        \insertion{(0.6,-0.8)}{};
        \projection{(0,0.8)}{j};
        \LeftModuleAction{(0,0)};
        \RightModuleAction{(0,0.4)};
        \draw[thick, decoration = {markings, mark=at position 0.2 with {\arrow[scale=1]{stealth}; \node[right, font=\scriptsize]{${}_g\Gamma$};}}, postaction=decorate] (0,-0.8) -- (0,0.8);
        \draw[algebra, thick, decoration = {markings, mark=at position 0.6 with {\arrow[scale=1]{stealth}; \node[left, font=\scriptsize]{$\hat{G_1}$};}}, postaction=decorate] (-0.6,-0.8) to[in=240, out=90] (-0.12,0);
        \draw[algebra, thick, decoration = {markings, mark=at position 0.6 with {\arrow[scale=1]{stealth}; \node[right, font=\scriptsize]{$\hat{G_1}$};}}, postaction=decorate] (0.6,-0.8) to[in=300, out=90] (0.12,0.4);
        \draw[simple] (-0.6,-1.2) -- (-0.6,-0.8);
        \draw[simple] (0,-1.2) -- (0,-0.8);
        \draw[simple] (0.6,-1.2) -- (0.6,-0.8);
        \draw[simple] (0,0.8) -- (0,1.2);
        \draw[](-0.8,-1.1) node[below]{\scriptsize$(g_2,1)$};
        \draw[](0,-1.1) node[below] {\scriptsize$(g,h)$};
        \draw[](0.8,-1.1) node[below] {\scriptsize$(g_1,1)$};
        \draw[](0,1.1) node[above] {\scriptsize$(g_2gg_1,h)$};
    \end{tikzpicture}
    }}
    = \Gamma^{-1}(g_2)_{ij}.
\end{equation}

We now invoke Theorems~7.12.16 and~7.10.1 of~\cite{gelakiTensorCategories2015}. Taken together, they imply that any module category over \({}_A\mathcal{C}_A\) is equivalent to one of the form \({}_A\mathcal{C}_B\), where \(A\) and \(B\) are algebra objects in \(\mathcal{C}\). Here we use the shorthand \({}_A\mathcal{C}_B\) for the category \(\bimod_{\mathcal{C}}(A,B)\) of \((A,B)\)-bimodules in \(\mathcal{C}\), and we will adopt the same convention throughout.

At this point it is helpful to distinguish two closely related notions of ``module'' that appear in the literature. The first is a \emph{module category} over a monoidal category; the second is a \emph{category of module objects} over an algebra object in a monoidal category. In our setting, Theorem~7.10.1 of~\cite{gelakiTensorCategories2015} provides the precise bridge between these viewpoints, allowing us to move freely between them. For clarity, we will refer to the action of \({}_A\mathcal{C}_A\) on \({}_A\mathcal{C}_B\) as the \emph{module product}, and to the underlying algebra action on individual objects as the \emph{module action}. Concretely, the module product is the functor
\begin{equation}\label{mod act}
    \begin{aligned}
        \odot: {}_A\mathcal{C}_A \times {}_A\mathcal{C}_B &\to {}_A\mathcal{C}_B ; \\
        M\odot N &\coloneqq M \otimes_A N,
    \end{aligned}
\end{equation}
where \(M\otimes_A N\) denotes the relative tensor product over \(A\). In particular, \(M\otimes_A N\) can be realized as the image of an idempotent endomorphism on \(M\otimes N\) (equivalently, as the coequalizer implementing the identification of the right \(A\)-action on \(M\) with the left \(A\)-action on \(N\)); in the graphical calculus used below, this idempotent is represented by the following morphism, which is straightforward to check is a projector:
\begin{equation}
    P_{\otimes_A}=
    \vcenter{\hbox{
    \begin{tikzpicture}
        \draw[algebra, thick, decoration = {markings, mark=at position 0.5 with {\arrow[scale=1]{stealth}; \node[right, font=\scriptsize]{$\hat{G_1}$};}}, postaction=decorate]
        (0,-0.25) to[in=300, out=180] (-1+0.12,0.5);
        \draw[algebra, thick, decoration = {markings, mark=at position 0.5 with {\arrow[scale=1]{stealth};}}, postaction=decorate]
        (0,-0.25) to[in=240, out=0] (1-0.12,0.5);
        \draw[algebra, thick, decoration = {markings, mark=at position 0.5 with {\arrow[scale=1]{stealth}}}] (0,-0.75) -- (0,-0.25);
        \draw[thick, decoration = {markings, mark=at position 0.5 with {\arrow[scale=1]{stealth}; \node[left, font=\scriptsize]{$M$};}}, postaction=decorate] (-1,-1) -- (-1,1);
        \draw[thick, decoration = {markings, mark=at position 0.5 with {\arrow[scale=1]{stealth}; \node[right, font=\scriptsize]{$N$};}}, postaction=decorate] (1,-1) -- (1,1);
        \RightModuleAction{(-1,0.5)};
        \LeftModuleAction{(1,0.5)};
        \multiplication{(0,-0.25)};
        \unit{(0,-0.75)};
    \end{tikzpicture}
    }}
\end{equation}

Therefore, specifying an \(\mathcal{H}\)-module structure on \(\Vecc\) is equivalent to specifying an algebra object \(\hat{H}\) in \(\mathcal{G}\) such that the resulting bimodule category satisfies
\[
{}_{\hat{G_1}}\!\mathcal{G}_{\hat{H}} \simeq \Vecc.
\]
Recall that the simple objects in \({}_{\hat{G_1}}\!\mathcal{G}\) can be described in terms of \(G\)-graded \(\hat{G}\)-algebras; more concretely, they are represented by
\[
M_g \coloneqq \bigoplus_{h\in G}(g,h),
\]
with the module structure induced by group multiplication. In order for \({}_{\hat{G_1}}\!\mathcal{G}_{\hat{H}}\) to collapse to \(\Vecc\), which has a unique simple object, the subgroup \(H\subseteq G\times G\) underlying \(\hat{H}\) must satisfy two basic constraints.

First, the projection of \(H\) onto the second factor must be surjective. If instead the second component ranges only over a proper subgroup \(N\subsetneq G\), then simples in \({}_{\hat{G_1}}\!\mathcal{G}_{\hat{H}}\) would be graded by right cosets \(G/N\), producing more than one simple object.

Second, the fiber of this projection must be trivial: for each \(h\in G\) there must be a unique \(g\in G\) such that \((g,h)\in H\). Equivalently, \(H\) cannot contain a nontrivial subgroup of the form \((N,1)\) with \(N\subseteq G\), since such a stabilizer would lead to additional simples labeled by irreducible representations of \(N\).

The only subgroups \(H\subseteq G\times G\) satisfying both conditions are graphs of endomorphisms,
\begin{equation}
    H_\phi \coloneqq \{\left(\phi(h),h\right)\mid h \in G\},
\end{equation}
with \(\phi\in\End(G)\).

There do exist other subgroup algebras of dimension larger than \(|G|\) that can also define bimodule categories; however, these necessarily involve twisted (in particular, anomalous) data. In our setting, such twists correspond to phases that carry an anomaly under \(G\) or \(\Rep(G)\) and hence are not intrinsically mixed. We defer a detailed discussion of this point to Sec.~\ref{discussion}.

To summarize, \(\mathcal{H}\)-module structures on \(\Vecc\) are parametrized (in the untwisted, anomaly-free sector relevant here) by endomorphisms \(\phi\in\End(G)\). For each \(\phi\), the corresponding module category is ${}_{\hat{H_\phi}}\mathcal{G}_{\hat{G_1}}$,

and it has a single simple object
\[
\Phi \coloneqq \bigoplus_{g,h}(g,h),
\]
whose left and right module actions are given by left and right group multiplication, respectively.

\subsection{fiber functor}
In this subsection we determine the explicit \emph{module product} of \(\mathcal{H}\) on the module category constructed in Sec.~\ref{Rep(G) times G-module structures on Vecc}, and then extract the associated fiber functor. Conceptually, this step connects the ``static'' classification of module categories (i.e., which \(\mathcal{H}\)-modules over \(\Vecc\) exist) to the \emph{monoidal data} that distinguish SPT phases: the associativity constraints of the module product precisely encode the tensor-compatibility isomorphisms of the fiber functor.

Concretely, for a module category \(M\simeq\Vecc\) over a monoidal category \(A\), the fiber functor \(F\) can be reconstructed from the action \(\odot\) by restricting to the distinguished simple object \(1\in M\). Namely,
\begin{equation}
    F(X) \equiv X\odot 1 \iff X \odot M \equiv F(X) \otimes M,\\
\end{equation}
and the coherence of the module associator translates into the monoidal structure of \(F\):
\begin{equation}\label{fiber functor}
    X \odot (Y \odot 1) \xrightarrow[m_{X,Y,1}]{\sim} (X \otimes Y) \odot 1\\
    \iff F(X) \otimes F(Y) \xrightarrow[J_{X,Y}]{\sim} F(X \otimes Y).
\end{equation}

Recall from Eq.~\ref{mod act} that the module product is implemented by the relative tensor product. In our realization \(\mathcal{H}\simeq {}_{\hat{G_1}}\!\mathcal{G}_{\hat{G_1}}\) acting on the module category \({}_{\hat{G_1}}\!\mathcal{G}_{\hat{H_\phi}}\), this action takes the form
\begin{align}
{}_{\hat{G_1}}\!\mathcal{G}_{\hat{G_1}} \times {}_{\hat{G_1}}\!\mathcal{G}_{\hat{H_\phi}} \xrightarrow{\otimes_{\hat{G_1}}} {}_{\hat{G_1}}\! \mathcal{G}_{\hat{H_\phi}}: \quad
{}_g\Gamma \otimes_{\hat{G_1}} \!\Phi \simeq \Phi^{\oplus d_\Gamma}.
\end{align}
Importantly, the above isomorphism is not canonical: while the object
\[
{}_g\Gamma \otimes_{\hat{G_1}}\! \Phi
\]
is isomorphic to a direct sum \(\Phi^{\oplus d_\Gamma}\) as an object of \({}_{\hat{G_1}}\!\mathcal{G}_{\hat{H_\phi}}\), the identification depends on a choice of intertwining data. This non-canonicity is precisely what gives rise to the monoidal structure maps \(J_{X,Y}\) in Eq.~\ref{fiber functor}, and hence to the eventual classification invariant valued in \(\Endo(G)/\Inn(G)\).

More concretely, the isomorphism in
\[
{}_g\Gamma \otimes_{\hat{G_1}}\! \Phi \simeq \Phi^{\oplus d_\Gamma}
\]
is not necessarily the identity map, because the relative tensor product inherits a nontrivial left \(\hat{G_1}\)-action from \({}_g\Gamma\). To make this explicit, define a basis of \({}_g\Gamma \otimes_{\hat{G_1}} \Phi\) as follows. Consider the image of
\(
\hat{e}(h'\bar{h},g)_i\times \phi(h,\bar{g}g')
\)
in the quotient defining \(\otimes_{\hat{G_1}}\), and denote the resulting basis vectors by \(\hat{\mu}(h',g')_i\) (here \(h\) is irrelevant modulo \(\hat{G_1}\)). With respect to this basis, the induced \((\hat{G_1},\hat{H_\phi})\)-bimodule action takes the form
\begin{equation}
    (h_1,e) \otimes \hat{\mu}(h,g)_i \otimes (\phi(h_2),h_2) \mapsto \sum_j\Gamma^{-1}(h_1)_{ij}\hat{\mu}(h_1h\phi(h_2),gh_2)_j,
\end{equation}
which mirrors the defining left action of \({}_g\Gamma\). In the next step, we will use a suitable change of basis to isolate \(d_\Gamma\) canonical copies of \(\Phi\); the resulting comparison maps between different tensor-product orderings will then determine \(J_{X,Y}\).
To trivialize the residual \(\Gamma\)-dependence in the left \(\hat{G_1}\)-action, it is convenient to perform a further change of basis. Define
\begin{equation}
    \hat{\mu'}(h,g)_i \coloneq \sum_j \Gamma(\phi(h)\bar{g})_{ij} \hat{\mu}(h,g)_j,
\end{equation}
which is an invertible transformation on the multiplicity index for each fixed \((h,g)\). A direct substitution into the bimodule action shows that the \(\Gamma^{-1}(h_1)\) factor is absorbed, and the action becomes purely regular:
\begin{equation}
    (h_1,e) \otimes \hat{\mu'}(h,g)_i \otimes (\phi(h_2),h_2) \mapsto \hat{\mu'}(h_1h\phi(h_2),gh_2)_i.
\end{equation}
In particular, for each fixed \(i\), the subspace
\[
\bigoplus_{g,h} \hat{\mu'}(h,g)_i
\]
is stable under the left \(\hat{G_1}\)- and right \(\hat{H_\phi}\)-actions, and hence forms a copy of \(\Phi\) as a \((\hat{G_1},\hat{H_\phi})\)-bimodule. Consequently, we obtain a decomposition
\[
{}_g\Gamma \otimes_{\hat{G_1}}\! \Phi \simeq \Phi^{\oplus d_\Gamma},
\]
where the \(d_\Gamma\) summands are precisely the subspaces labeled by \(i\). The remaining ambiguity in identifying these summands across iterated tensor products is what will be captured by the coherence data summarized in the following diagram (with \(\bigoplus \equiv \bigoplus_{g,h}\)):
\[\begin{tikzcd}
	\Phi & {\bigoplus \hat{\mu_1}(h,g_1g)^{\oplus d_{\Gamma_1}}} & {\Phi^{\oplus d_{\Gamma_1}}} \\
	\\
	& {\bigoplus \big( \hat{\mu_2}(h,g_2g_1g)^{\oplus d_{\Gamma_2}} \otimes \hat{\mu_1}(h,g_2g_1g)^{\oplus d_{\Gamma_1}}\big)} & {\bigoplus \hat{\mu_2}(h,g_2g_1g)^{\oplus d_{\Gamma_2} d_{\Gamma_1}}} \\
	\\
	&& {\Phi^{\oplus d_{\Gamma_1} d_{\Gamma_2}}}
	\arrow["{{{{{}_{g_1}\Gamma_1 \otimes_{\hat{G_1}}}}}}"{description}, from=1-1, to=1-2]
	\arrow["{{{{{}_{g_2g_1}(\Gamma_2 \otimes \Gamma_1) \otimes_{\hat{G_1}}}}}}"', from=1-1, to=3-2]
	\arrow["{{{{\bigoplus \Gamma_1(\phi(g_1g)\bar{h})}}}}"{description}, from=1-2, to=1-3]
	\arrow["{{\Gamma_1(\phi(g_2))}}"{description},  Rightarrow, from=1-3, to=3-2, shorten >=20pt, shorten <=20pt]
	\arrow["{{{{{}_{g_2}\Gamma_2 \otimes_{\hat{G_1}}}}}}"{description}, from=1-3, to=3-3]
	\arrow["{{{{\bigoplus \big( (\Gamma_2 \otimes \Gamma_1)(\phi(g_2g_1g)\bar{h}) \big)}}}}"', from=3-2, to=5-3]
	\arrow["{{{{\bigoplus \big(\Gamma_2(\phi(g_2g_1g)\bar{h}) \otimes I_{d_{\Gamma_1}}\big)}}}}"{description}, from=3-3, to=5-3]
\end{tikzcd},\]
Recalling Eq.~\ref{fiber functor}, the comparison between the two natural ways of identifying
\(
{}_{g_2}\Gamma_2 \odot ({}_{g_1}\Gamma_1 \odot 1)
\)
and
\(
({}_{g_2}\Gamma_2 \otimes {}_{g_1}\Gamma_1)\odot 1
\)
yields the monoidal structure isomorphism
\begin{equation}
    J_{{}_{g_2}\Gamma_2, {}_{g_1}\Gamma_1}=\Gamma_1(\phi(g_2)).
\end{equation}
Equivalently, the nontrivial associativity data of the module product is encoded by evaluating the representation \(\Gamma_1\) on the \(\phi\)-image of the group element \(g_2\). This is the precise sense in which the endomorphism \(\phi\) controls the ``mixed'' coupling between the \(\Rep(G)\) and \(G\) sectors.

It follows that the fiber functor is characterized, up to equivalence, by \(\phi\) modulo inner automorphisms, namely by a class in \(\Endo(G)/\Inn(G)\). Indeed, if \(f\in\Inn(G)\) is an inner automorphism, then for every representation \(\Gamma\) the representations \(\Gamma\circ f\) and \(\Gamma\) are equivalent, so replacing \(\phi\) by \(f\circ \phi\) does not change the resulting \(J\)-symbols up to a natural monoidal isomorphism. Consequently, only the outer part of \(\phi\) can label distinct \(\mathcal{H}\)-SPT phases.

To summarize, the classification data of the intrinsically mixed \(\Rep(G)\times G\) SPT phases constructed here can be packaged into the monoidal structure \(J_{X,Y}\) of the associated fiber functor, and this structure is completely determined by the class of \(\phi\) in \(\Endo(G)/\Inn(G)\). From the physical viewpoint, the same \(J\)-symbols provide the topological response: they dictate how symmetry defects fuse in the effective \((1+1)\)D theory, equivalently the phase factors (or matrix intertwiners) acquired when a \(G\)-defect is moved past a \(\Rep(G)\)-defect. In the lattice realization, this response manifests as the \(\phi\)-twisted action across the domain wall and, upon dimensional reduction, as the protected edge-mode transformation laws encoded by the boundary symmetry operators.

\section{Characterize SPT by condensable algebras}\label{Characterize SPT by condensable algebras}
Having classified the \(\Rep(G)\times G\) SPT phases in terms of fiber functors (equivalently, module categories over \(\mathcal{H}\) with underlying category \(\Vecc\)), we now relate this abstract classification to the more physical language of anyon condensation and domain walls in the corresponding \((2+1)\)D topological order. Concretely, we would like to identify the condensable (Lagrangian) algebra in the bulk that realizes each SPT phase, and to match it with the boundary data encoded by the \(\mathcal{H}\)-module structure. This step provides a bridge between the categorical invariant extracted in the previous subsection (the monoidal data \(J_{X,Y}\), hence \(\phi\in\Endo(G)/\Inn(G)\)) and the lattice/domain-wall picture developed later in Sec.~\ref{Interpretation in physics}.

Within the SymTFT framework, which we will review in detail in Sec.~\ref{Interpretation in physics}, every \(1+1d\) gapped phase with a categorical symmetry \(\mathcal{C}\) corresponds one-to-one with a physical gapped boundary in a sandwich construction. In this construction, the bulk topological order is associated with the topological skeleton \(\Zc(\mathcal{C})\). The categorical description of such a physical boundary has a ``double identity'': it can be viewed both as a bimodule category over an algebra \(A\) in \(\mathcal{C}\), and as a right module category over a Lagrangian algebra \(\mathcal{A}\) in \(\Zc(\mathcal{C})\). Our task in this section is to identify these two algebras. In physical terms, these correspond respectively to being ``gauged by \(A\)'' and ``condensed by \(\mathcal{A}\)''.

For a symmetry-protected topological phase, no nontrivial order parameters (i.e., anyon lines) can stretch between the two boundaries. Categorically, this is equivalent to requiring that the domain wall between \(\mathcal{C}\) and \({}_A\mathcal{C}_A\), namely \(\mathcal{C}_A\), is trivial (vacuum). This is precisely why categorical \(1+1d\) SPT phases are classified by fiber functors.

In our case, where \(\mathcal{C}=\mathcal{H}\), we therefore seek algebra objects \(A\in\mathcal{H}\) and \(\mathcal{A}_\phi\in \Zc(\mathcal{H})\simeq \mathcal{D}_G \boxtimes \mathcal{D}_G\) such that
\begin{align}
    \mathcal{H}_A \simeq {}_{\hat{G_1}}\!\mathcal{G}_{\hat{H_\phi}},\label{A}\\
    (\mathcal{D}_G \boxtimes \mathcal{D}_G)_{\mathcal{A}_\phi} \simeq {}_A \mathcal{H}_A.\label{large A}
\end{align}

Let us first focus on Eq.~\ref{A}. To this end, we briefly recall the notion of the internal hom. For a left \(\mathcal{C}\)-module category \(\mathcal{M}\) and objects \(X,Y\in\mathcal{M}\), the internal hom \(\Inhom_{\mathcal{C}}(X,Y)\in\mathcal{C}\) is characterized (when it exists) by the natural isomorphism
\begin{equation*}
    \alpha:\hom_{\mathcal{M}}(\bullet \odot X, Y) \simeq \hom_{\mathcal{C}}(\bullet, \Inhom_{\mathcal{C}}(X,Y)).
\end{equation*}
We denote by \(\ev_{X,Y}:\Inhom_{\mathcal{C}}(X,Y)\odot X\to Y\) the morphism corresponding, under \(\alpha\), to \(\id_{\Inhom_{\mathcal{C}}(X,Y)}\); equivalently, \(\ev_{X,Y}\) is the evaluation map implementing the universal property of \(\Inhom_{\mathcal{C}}(X,Y)\).

Now specialize to \(\mathcal{C}=\mathcal{H}\) and \(\mathcal{M}={}_{\hat{G_1}}\!\mathcal{G}_{\hat{H_\phi}}\). Applying Theorem~7.10.1 of~\cite{gelakiTensorCategories2015}, we obtain the algebra object
\begin{equation}
    A=\Inhom_{\mathcal{H}}(G_1\otimes H_\phi, G_1\otimes H_\phi).
\end{equation}
Moreover, \(A\) carries a canonical algebra structure induced by composition of internal endomorphisms. Concretely, for any \(X\in\mathcal{M}\), the multiplication on \(\Inhom_{\mathcal{H}}(X,X)\) is determined by the composite (suppressing identity morphisms)
\begin{multline}\label{internal algebra}
    \Inhom_{\mathcal{H}}(X,X) \otimes \Inhom_{\mathcal{H}}(X,X) \odot X \xrightarrow{\ev_{X,X} \circ \ev_{X,X}} X\\
    \xmapsto{\alpha_X} \Inhom_{\mathcal{H}}(X,X) \otimes \Inhom_{\mathcal{H}}(X,X) \xrightarrow{m} \Inhom_{\mathcal{H}}(X,X),  
\end{multline}
which is the internal version of the usual multiplication by composition of endomorphisms.

Using the graphical calculus, we next observe the following diagrammatic isotopy, which implements the universal property of the relative tensor product and amounts to ``sliding'' the \(\hat{G_1}\)-actions through the pairing. In particular, for \(X,Y,Z\in {}_{\hat{G_1}}\!\mathcal{G}_{\hat{H_\phi}}\) one has
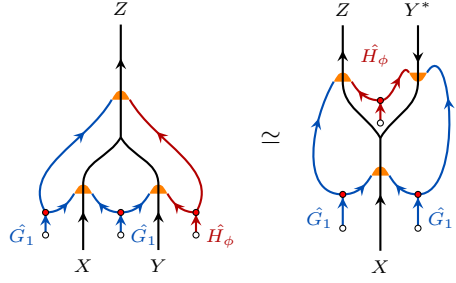
\begin{multline}
    \hom_{{}_{\hat{G_1}}\!\mathcal{G}_{\hat{H_\phi}}}(X \otimes_{{\hat{G_1}}} Y, Z)=\\
    \vcenter{\hbox{
    \begin{tikzpicture}
        \LeftModuleAction{(0,0.5)};
        \LeftModuleAction{(-0.5,-0.75)};
        \LeftModuleAction{(0.5,-0.75)};
        \RightModuleAction{(0,0.5)};
        \RightModuleAction{(-0.5,-0.75)};
        \RightModuleAction{(0.5,-0.75)};
        \draw[thick, decoration = {markings, mark=at position 0.7 with {\arrow[scale=1]{stealth}};}, postaction=decorate] (0,0) -- (0,1.5);
        \draw[thick, decoration = {markings, mark=at position 0.3 with {\arrow[scale=1]{stealth}};}, postaction=decorate] (-0.5,-1.5) -- (-0.5,-0.75+0.12) to[in=240, out=90] (0,0);
        \draw[thick, decoration = {markings, mark=at position 0.3 with {\arrow[scale=1]{stealth}};}, postaction=decorate] (0.5,-1.5) -- (0.5,-0.75+0.12) to[in=300, out=90] (0,0);
        \draw[thick, algebra, decoration = {markings, mark=at position 0.7 with {\arrow[scale=1]{stealth}};}, postaction=decorate] (0,-1) to[in=300, out=180] (-0.5+0.12,-0.75);
        \draw[thick, algebra, decoration = {markings, mark=at position 0.7 with {\arrow[scale=1]{stealth}};}, postaction=decorate] (0,-1) to[in=240, out=0] (0.5-0.12,-0.75);
        \draw[thick, algebra, decoration = {markings, mark=at position 0.9 with {\arrow[scale=1]{stealth}};{\node[below right, font=\scriptsize]{$\hat{G_1}$};}}, postaction=decorate] (0,-1.3) -- (0,-1);
        \unit{(0,-1.3)};
        \multiplication{(0,-1)};
        \draw[thick, algebra, decoration = {markings, mark=at position 0.9 with {\arrow[scale=1]{stealth}};{\node[below left, font=\scriptsize]{$\hat{G_1}$};}}, postaction=decorate] (-1,-1.3) -- (-1,-1);
        \draw[thick, algebra, decoration = {markings, mark=at position 0.7 with {\arrow[scale=1]{stealth}};}, postaction=decorate] (-1,-1) to[in=240, out=0] (-0.5-0.12,-0.75);
        \draw[thick, algebra, decoration = {markings, mark=at position 0.7 with {\arrow[scale=1]{stealth}};}, postaction=decorate] (-1,-1) to[in=240, out=120] (-0.12,0.5);
        \unit{(-1,-1.3)};
        \multiplication{(-1,-1)};
        \draw[thick, algebra2, decoration = {markings, mark=at position 0.9 with {\arrow[scale=1]{stealth}; \node[below right, font=\scriptsize]{$\hat{H_\phi}$};}}, postaction=decorate] (1,-1.3) -- (1,-1);
        \draw[thick, algebra2, decoration = {markings, mark=at position 0.7 with {\arrow[scale=1]{stealth}};}, postaction=decorate] (1,-1) to[in=300, out=180] (0.5+0.12,-0.75);
        \draw[thick, algebra2, decoration = {markings, mark=at position 0.7 with {\arrow[scale=1]{stealth}};}, postaction=decorate] (1,-1) to[in=300, out=60] (0.12,0.5);
        \unit{(1,-1.3)};
        \multiplication{(1,-1)};
        \draw[] (0.5,-1.5) node[below]{\scriptsize$Y$};
        \draw[] (-0.5,-1.5) node[below]{\scriptsize$X$};
        \draw[] (0,1.5) node[above]{\scriptsize$Z$};
    \end{tikzpicture}
    }}
    \simeq\vcenter{\hbox{
    \begin{tikzpicture}
        \begin{scope}[rotate around={180:(0.5,0.75+0.06)}]
            \LeftModuleAction{(0.5,0.75)};
            \RightModuleAction{(0.5,0.75)};
        \end{scope}
        \LeftModuleAction{(-0.5,0.75)};
        \RightModuleAction{(-0.5,0.75)};
        \LeftModuleAction{(0,-0.5)};
        \RightModuleAction{(0,-0.5)};
        \draw[thick, decoration = {markings, mark=at position 0.5 with {\arrow[scale=1]{stealth}};}, postaction=decorate] (0,-1.5) -- (0,0);
        \draw[thick, decoration = {markings, mark=at position 0.3 with {\arrow[scale=1]{stealth}};}, postaction=decorate] (0.5,1.5) -- (0.5,0.75) to[in=60, out=270] (0,0);
        \draw[thick, decoration = {markings, mark=at position 0.8 with {\arrow[scale=1]{stealth}};}, postaction=decorate] (0,0) to[in=270, out=120] (-0.5,0.75) -- (-0.5,1.5);
        \draw[thick, algebra2, decoration = {markings, mark=at position 0.9 with {\arrow[scale=1]{stealth}};}, postaction=decorate] (0,0.2) -- (0,0.5);
        \draw[thick, algebra2, decoration = {markings, mark=at position 0.7 with {\arrow[scale=1]{stealth}};}, postaction=decorate] (0,0.5) to[in=300, out=180] (-0.5+0.12,0.75);
        \draw[thick, algebra2, decoration = {markings, mark=at position 0.7 with {\arrow[scale=1]{stealth}; \node[above left, font=\scriptsize]{$\hat{H_\phi}$};}}, postaction=decorate] (0,0.5) to[in=120, out=0] (0.5-0.12,0.75+0.12);
        \multiplication{(0,0.5)};
        \unit{(0,0.2)};
        \draw[thick, algebra, decoration = {markings, mark=at position 0.9 with {\arrow[scale=1]{stealth}; \node[below right, font=\scriptsize]{$\hat{G_1}$};}}, postaction=decorate] (0.5,-1.2) -- (0.5,-0.75);
        \draw[thick, algebra, decoration = {markings, mark=at position 0.7 with {\arrow[scale=1]{stealth}};}, postaction=decorate] (0.5,-0.75) to[in=300, out=180] (0.12,-0.5);
        \draw[thick, algebra, decoration = {markings, mark=at position 0.7 with {\arrow[scale=1]{stealth}};}, postaction=decorate] (0.5,-0.75) to[in=60, out=0] (0.5+0.12,0.75+0.12);
        \multiplication{(0.5,-0.75)};
        \unit{(0.5,-1.2)};
        \draw[thick, algebra, decoration = {markings, mark=at position 0.9 with {\arrow[scale=1]{stealth}; \node[below left, font=\scriptsize]{$\hat{G_1}$};}}, postaction=decorate] (-0.5,-1.2) -- (-0.5,-0.75);
        \draw[thick, algebra, decoration = {markings, mark=at position 0.7 with {\arrow[scale=1]{stealth}};}, postaction=decorate] (-0.5,-0.75) to[in=240, out=0] (-0.12,-0.5);
        \draw[thick, algebra, decoration = {markings, mark=at position 0.7 with {\arrow[scale=1]{stealth};}}, postaction=decorate] (-0.5,-0.75) to[in=240, out=180] (-0.5-0.12,0.75);
        \draw[] (0,-1.5) node[below]{\scriptsize $X$};
        \draw[] (-0.5,1.5) node[above]{\scriptsize $Z$};
        \draw[] (0.5,1.5) node[above]{\scriptsize $Y^*$};
        \multiplication{(-0.5,-0.75)};
        \unit{(-0.5,-1.2)};
        \end{tikzpicture}
    }}
    \\=\hom_{\mathcal{H}}(X, Z \otimes_{\hat{H_\phi}} Y^*),
\end{multline}
where \(Y^*\) denotes the dual of \(Y\) as a right \(\hat{H_\phi}\)-module. In particular, comparing the two expressions shows that the internal hom in \(\mathcal{H}\) is realized by
\[
\operatorname{hom}_{\mathcal{H}}(Y,Z) = Z \otimes_{\hat{H_\phi}} Y^*.
\]
Applying this to \(X=Y=Z=G_1\otimes H_\phi\), we obtain
\[
A = G_1 \otimes H_\phi \otimes_{\hat{H_\phi}} H_\phi^* \otimes G_1^* = G_1 \otimes H_\phi^* \otimes G_1^*.
\]
Finally, Eq.~\ref{internal algebra} identifies the multiplication on \(A\) as the composition of internal endomorphisms; in the present realization this reduces to
\[
A \otimes A \xrightarrow{m_{H_\phi} \circ \operatorname{ev}_{G_1}} A
\]
(again suppressing identity morphisms), where \(m_{H_\phi}\) is the group multiplication of \(H_\phi\).

Next, we proceed to determine the corresponding Lagrangian algebra \(\mathcal{A}_\phi\). Before doing so, we note a useful simplification. Since \(A\) is an algebra object in
\(
\mathcal{H} = {}_{\hat{G_1}}\!\mathcal{G}_{\hat{G_1}},
\)
the bimodule category \({}_A\mathcal{H}_A\) can be canonically identified with \({}_A\mathcal{G}_A\). This identification follows from two complementary observations. 
On the one hand, given any object in \({}_A\mathcal{G}_A\), we may equip it with an \(\hat{G_1}\)-bimodule structure by restricting the \(A\)-actions along the unit embedding \(\hat{G_1} \xrightarrow{u} A\). The resulting left and right \(\hat{G_1}\)-actions are automatically compatible with the original \(A\)-bimodule structure, so the object may equally well be regarded as an object of \({}_A\mathcal{H}_A\). 
On the other hand, any object of \({}_A\mathcal{H}_A\) admits a canonical image in \({}_A\mathcal{G}_A\) under the forgetful functor that discards the \(\hat{G_1}\)-bimodule structure while retaining the underlying \(A\)-bimodule in \(\mathcal{G}\). These two constructions are inverse to each other up to canonical isomorphism, yielding the claimed identification \({}_A\mathcal{H}_A \simeq {}_A\mathcal{G}_A\).
As a result, it is technically more convenient to work with \({}_A\mathcal{G}_A\) when computing \(\mathcal{A}_\phi\), rather than analyzing \({}_A\mathcal{H}_A\) directly.

It's known that~\cite{gelakiTensorCategories2015}, the Drinfeld centers of \(\mathcal{C}\) and \({}_A\mathcal{C}_A\) are equivalent in a canonical way. More precisely, there is a braided equivalence
\begin{equation}\label{center eq}
    \Zc(\mathcal{C}) \xrightarrow{\simeq} \Zc({}_A\mathcal{C}_A): \quad Z \mapsto Z \otimes A, 
\end{equation}
which sends an object \(Z\in\Zc(\mathcal{C})\) (equipped with its half-braiding) to the object \(Z\otimes A\) viewed in \(\Zc({}_A\mathcal{C}_A)\).

Under this equivalence, the \((A,A)\)-bimodule structure on \(Z\otimes A\) is specified as follows. The right \(A\)-action is the obvious one on the last tensor factor, while the left \(A\)-action uses the half-braiding of \(Z\). Concretely, the left action is given by the composite
\begin{equation}
    A \otimes(Z\otimes A) \xrightarrow{\gamma_{A,Z} \otimes \id_A} Z \otimes A \otimes A \xrightarrow{\id_Z \otimes m} Z \otimes A,
\end{equation}
where \(\gamma_{A,Z}\) denotes the half-braiding isomorphism associated to \(Z\) and \(m\) is the multiplication of \(A\). In the graphical calculus, these left and right module structures are depicted by:
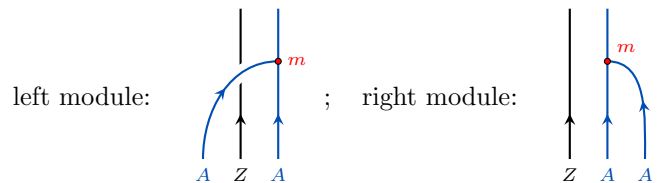
\begin{equation*}
    \text{left module:} \quad
    \vcenter{\hbox{
    \begin{tikzpicture}
        \draw[thick, decoration = {markings, mark=at position 0.3 with {\arrow[scale=1]{stealth}};}, postaction=decorate] (0,-1) -- (0,1);
        \draw [algebra, preaction={draw=white,line width=6pt}, thick, decoration = {markings, mark=at position 0.55 with {\arrow[scale=1]{stealth}}}, postaction=decorate] (-0.5,-1) to[in=180, out=90] (0.5,0.3);
        \draw[algebra, thick, decoration = {markings, mark=at position 0.3 with {\arrow[scale=1]{stealth}};}, postaction=decorate] (0.5,-1) -- (0.5,1);
        \multiplication{(0.5,0.3)};
        \draw[red] (0.5,0.3) node[right] {\scriptsize $m$};
        \draw[] (0,-1) node[below]{\scriptsize $Z$};
        \draw[algebra] (0.5,-1) node[below]{\scriptsize $A$};
        \draw[algebra] (-0.5,-1) node[below]{\scriptsize $A$};
    \end{tikzpicture}
    }}; \quad
    \text{right module:} \quad
    \vcenter{\hbox{
    \begin{tikzpicture}
        \draw[thick, decoration = {markings, mark=at position 0.3 with {\arrow[scale=1]{stealth}};}, postaction=decorate] (0,-1) -- (0,1);
        \draw [algebra, preaction={draw=white,line width=6pt}, thick, decoration = {markings, mark=at position 0.5 with {\arrow[scale=1]{stealth}}}, postaction=decorate] (1,-1) to[in=0, out=90] (0.5,0.3);
        \draw[algebra, thick, decoration = {markings, mark=at position 0.3 with {\arrow[scale=1]{stealth}};}, postaction=decorate] (0.5,-1) -- (0.5,1);
        \multiplication{(0.5,0.3)};
        \draw[red] (0.5,0.3) node[above right] {\scriptsize $m$};
        \draw[algebra] (0.5,-1) node[below]{\scriptsize $A$};
        \draw[algebra] (1,-1) node[below]{\scriptsize $A$};
        \draw[] (0,-1) node[below]{\scriptsize $Z$};
    \end{tikzpicture}
    }}.
\end{equation*}
It is also well known that the boundary topological order \(\mathcal{C}\) can be recovered from its bulk \(\Zc(\mathcal{C})\) by condensing the canonical Lagrangian algebra \(I_{\mathcal{C}}(\mathbf{1})\). Equivalently,
\[
\mathcal{C} \simeq \Zc(\mathcal{C})_{I_{\mathcal{C}}(\mathbf{1})}.
\]
Here \(I_{\mathcal{C}}\) denotes the right adjoint of the forgetful functor \(F:\Zc(\mathcal{C})\to \mathcal{C}\), characterized by the adjunction isomorphism
\begin{equation}\label{adjoint}
     \hom_{\Zc(\mathcal{C})}(Z,I_{\mathcal{C}}(c)) \simeq \hom_{\mathcal{C}}(F(Z),c).
\end{equation}

We now return to our setting. Since the unit object \(\mathbf{1}_{{}_A\mathcal{G}_A}\) of the bimodule category \({}_A\mathcal{G}_A\) is simply \(A\) itself, Eqs.~\ref{center eq} and~\ref{adjoint} together imply that the corresponding Lagrangian algebra \(\mathcal{A}_\phi\in \mathcal{O}(\mathcal{D}_G^2)\) decomposes as
\begin{equation}
    \mathcal{A}_\phi = \bigoplus_{Z \in \mathcal{O}(\mathcal{D}_G^2)} Z^{\oplus \dim \left(\hom_{{}_A\mathcal{G}_A}(Z\otimes A, A)\right)}.
\end{equation}
Thus, determining \(\mathcal{A}_\phi\) reduces to computing the multiplicity space of \((A,A)\)-bimodule morphisms from \(Z\otimes A\) to \(A\). In other words, we must characterize all bimodule maps \(F: Z\otimes A\to A\); such an \(F\) is constrained by the requirement that it intertwines both the left and right \(A\)-actions, i.e., it must satisfy:
\begin{equation}\label{F condition}
    \vcenter{\hbox{
    \begin{tikzpicture}
        \draw[thick, decoration = {markings, mark=at position 0.3 with {\arrow[scale=1]{stealth}};}, postaction=decorate] (0,-1) to[in=180, out=90] (0.5,0);
        \draw [algebra, preaction={draw=white,line width=6pt}, thick, decoration = {markings, mark=at position 0.55 with {\arrow[scale=1]{stealth}}}, postaction=decorate] (1,-1) to[in=0, out=90] (0.5,0.5);
        \draw[algebra, thick, decoration = {markings, mark=at position 0.3 with {\arrow[scale=1]{stealth}};}, postaction=decorate] (0.5,-1) -- (0.5,1);
        \multiplication{(0.5,0.5)};
        \draw[red] (0.5,0.5) node[above right]{\scriptsize $m$};
        \fill[cyan] (0.5,0) circle(0.04);
        \draw[] (0.5,0) node[above left]{\scriptsize $F$};
        \draw[] (0.5,0) circle(0.04);
        \draw[algebra] (0.5,-1) node[below]{\scriptsize $A$};
        \draw[algebra] (1,-1) node[below]{\scriptsize $A$};
        \draw[] (0,-1) node[below]{\scriptsize $Z$};
    \end{tikzpicture}
    }}
    =\vcenter{\hbox{
    \begin{tikzpicture}
        \draw[thick, decoration = {markings, mark=at position 0.3 with {\arrow[scale=1]{stealth}};}, postaction=decorate] (0,-1) to[in=180, out=90] (0.5,0.5);
        \draw [algebra, preaction={draw=white,line width=6pt}, thick, decoration = {markings, mark=at position 0.55 with {\arrow[scale=1]{stealth}}}, postaction=decorate] (1,-1) to[in=0, out=90] (0.5,0);
        \draw[algebra, thick, decoration = {markings, mark=at position 0.3 with {\arrow[scale=1]{stealth}};}, postaction=decorate] (0.5,-1) -- (0.5,1);
        \multiplication{(0.5,0)};
        \draw[red] (0.5,0) node[left]{\scriptsize $m$};
        \fill[cyan] (0.5,0.5) circle(0.04);
        \draw[] (0.5,0.5) node[right]{\scriptsize $F$};
        \draw[] (0.5,0.5) circle(0.04);
        \draw[algebra] (0.5,-1) node[below]{\scriptsize $A$};
        \draw[algebra] (1,-1) node[below]{\scriptsize $A$};
        \draw[] (0,-1) node[below]{\scriptsize $Z$};
    \end{tikzpicture}
    }}
    \quad \text{and} \quad
    \vcenter{\hbox{
    \begin{tikzpicture}
        \draw[thick, decoration = {markings, mark=at position 0.3 with {\arrow[scale=1]{stealth}};}, postaction=decorate] (0,-1) to[in=180, out=90] (0.5,0.5);
        \draw [algebra, preaction={draw=white,line width=6pt}, thick, decoration = {markings, mark=at position 0.55 with {\arrow[scale=1]{stealth}}}, postaction=decorate] (-0.5,-1) to[in=180, out=90] (0.5,0);
        \draw[algebra, thick, decoration = {markings, mark=at position 0.3 with {\arrow[scale=1]{stealth}};}, postaction=decorate] (0.5,-1) -- (0.5,1);
        \multiplication{(0.5,0)};
        \draw[red] (0.5,0) node[right] {\scriptsize $m$};
        \fill[cyan] (0.5,0.5) circle(0.04);
        \draw[] (0.5,0.5) circle(0.04);
        \draw[] (0.5,0.5) node[right]{\scriptsize $F$};
        \draw[] (0,-1) node[below]{\scriptsize $Z$};
        \draw[algebra] (0.5,-1) node[below]{\scriptsize $A$};
        \draw[algebra] (-0.5,-1) node[below]{\scriptsize $A$};
    \end{tikzpicture}
    }}
    =\vcenter{\hbox{
    \begin{tikzpicture}
        \draw[thick, decoration = {markings, mark=at position 0.3 with {\arrow[scale=1]{stealth}};}, postaction=decorate] (0,-1) to[in=180, out=90] (0.5,0);
        \draw [algebra, preaction={draw=white,line width=6pt}, thick, decoration = {markings, mark=at position 0.55 with {\arrow[scale=1]{stealth}}}, postaction=decorate] (-0.5,-1) to[in=180, out=90] (0.5,0.5);
        \draw[algebra, thick, decoration = {markings, mark=at position 0.3 with {\arrow[scale=1]{stealth}};}, postaction=decorate] (0.5,-1) -- (0.5,1);
        \multiplication{(0.5,0.5)};
        \draw[red] (0.5,0.5) node[right] {\scriptsize $m$};
        \fill[cyan] (0.5,0) circle(0.04);
        \draw[] (0.5,0) circle(0.04);
        \draw[] (0.5,0) node[right]{\scriptsize $F$};
        \draw[] (0,-1) node[below]{\scriptsize $Z$};
        \draw[algebra] (0.5,-1) node[below]{\scriptsize $A$};
        \draw[algebra] (-0.5,-1) node[below]{\scriptsize $A$};
    \end{tikzpicture}
    }}.
\end{equation}

We now expand the bimodule constraints in Eq.~\ref{F condition} in components. Before doing so, we fix some notation. Recall that
\[
A=G_1 \otimes H_\phi^* \otimes G_1^*.
\]
We will write \(A\) in terms of components
\[
A(g_1,h_1,g_2) \coloneq (g_1,e)\otimes(\phi(h_1),h_1) \otimes (\bar{g_2},e),
\]
so that the multiplication induced by Eq.~\ref{internal algebra} takes the form
\[
A(g_1,h_1,g_2) \otimes A(g_3,h_2,g_4) \equiv \delta_{\bar{g_2}g_3}A(g_1,h_1h_2,g_4).
\]
In addition, we denote an irreducible object of \(\mathcal{D}_G^2\) by an anyon pair \(\left([g],\rho\right) \boxtimes ([h], \pi)\), where \([g]\) and \([h]\) are conjugacy classes and \(\rho\) and \(\pi\) are irreducible representations of the centralizers of \(g\) and \(h\), respectively. We write its basis components as \(\hat{e}(g,h)_{i,j}\), with \(i\) and \(j\) ranging over the dimensions of \(\rho\) and \(\pi\).

A first immediate consequence of Eq.~\ref{F condition} is that, whenever \(F\) is nonzero on a simple tensor \(\hat{e}(g,h)_{i,j}\otimes A(g_1,h_1,g_2)\), its image must be compatible with the delta function in the multiplication of \(A\). In particular, \(F\left(\hat{e}(g,h)_{i,j}\otimes A(g_1,h_1,g_2)\right)\) must be proportional to \(A(g_1,hh_1,g_2)\). Equating the corresponding group labels forces
\[
gg_1\phi(h_1)g_2=g_1\phi(hh_1)g_2,
\]
otherwise the image under \(F\) is necessarily zero. This constraint implies
\[
g=g_1\phi(h)\bar{g_1},
\]
and hence \([g]=[\phi(h)]\).

Accordingly, we may assume without loss of generality that \(F\) has the following component form:
\begin{equation}
    F\left(\hat{e}(g_1\phi(h)\bar{g_1},h)_{i,j}\otimes A(g_1,h_1,g_2)\right)=f_{i,j}(h,g_1,h_1,g_2) A(g_1,hh_1,\bar{g_2}).
\end{equation}

From the first equation in Eq.~\ref{F condition}, we deduce the constraint
\begin{equation}
    f_{i,j}(h,g_1,h_1,g_2)=f_{i,j}(h,g_1,h_1h_2,g_4), \quad \forall h_2,g_4, 
\end{equation}
which immediately implies that \(f_{i,j}\) is independent of \(h_1\) and \(g_2\). We may therefore write \(f_{i,j}(h,g_1,h_1,g_2)=f_{i,j}(h,g_1)\). Substituting this simplification into the second equation of Eq.~\ref{F condition} yields
\begin{equation}\label{F eq}
    F\left(\gamma\left(A(g_1,h_1,\bar{g_2}) \otimes \hat{e}(g_2\phi(h)\bar{g_2},h)_{i,j}\right) \otimes_m A(g_2,h_2,\bar{g_3})\right)
    =f_{i,j}(h,g_2) A(g_1,h_1hh_2,\bar{g_3}),
\end{equation}
where \(\gamma\) denotes the half-braiding and \(\otimes_m\) indicates that the tensor product on the right is followed by the multiplication of \(A\).

To analyze Eq.~\ref{F eq}, fix representatives \(c\) and \(d\) of the conjugacy classes \([h]\) and \([\phi(h)]\), respectively. Choose sets of elements \(\{p\}\) and \(\{q\}\) such that \(pc\bar{p}\) (respectively \(qd\bar{q}\)) runs through all elements of \([c]\) (respectively \([d]\)) exactly once. Let \(Y=\{y\}\) and \(Z=\{z\}\) denote the corresponding centralizers. Without loss of generality, we adopt the parametrization
\begin{equation}\label{convention}
    \phi(c) = d, \quad h = p c \bar{p}, \quad h_1 = p_1 y_1\bar{p}, \quad g_2 = q_2 z_2 \phi(\bar{p}), \quad g_1 \phi(p_1 y_1) = q_1 z_1,
\end{equation}
which organizes the variables appearing in Eq.~\ref{F eq} into conjugacy-class data and centralizer data. In these variables, Eq.~\ref{F eq} reduces to the component relation
\begin{equation}\label{f eq}
    \rho^{\dagger}(z_1 \bar{z_2})_{ik} \pi^{\dagger}(y_1)_{jl} f_{k,l}\left(p_1 c \bar{p_1}, q_1 z_1 \phi(\bar{y_1} \bar{p_1})\right) = f_{i,j}(p c \bar{p}, q_2 z_2 \phi(\bar{p})),
\end{equation}
valid for arbitrary \(z_1,z_2,y_1,p,p_1,q_1,q_2\).

Eliminating variables that play no essential role, we find that
\[
f_{i,j}(pc\bar{p},qz\phi(\bar{p}))=f_{i,j}(c,z)
\]
for any \(p,q\) and \(z\). Substituting this back, the condition further simplifies to
\begin{align}
    \rho(z_2 \bar{z_1}) \cdot f(c,z_1 \phi(\bar{y})) \cdot \bar{\pi}(y) = f(c,z_2),
\end{align}
where \(\bar{\pi}\) denotes the complex conjugate representation of \(\pi\). This equation shows that the existence of a nonzero solution \(f\) is equivalent to the existence of an intertwiner between \(\rho \circ \phi\) and \(\bar{\pi}\). Equivalently, since \(\phi(y)\in Z\) for \(y\in Y\), \(\phi\) induces the pullback representation \(\phi^*\rho(y)\coloneq \rho(\phi(y))\), and the intertwining condition may be stated as \(\phi^*\rho \to \bar{\pi}\) in \(\Rep(Y)\).

Consequently, the multiplicity of the anyon pair \(([\phi(h)],\rho)\boxtimes([h],\pi)\) in \(\mathcal{A}_\phi\) is given by the intertwining number between $\phi^*\rho$ and $\bar{\pi}$, i.e.,
\begin{equation}\label{pair}
    \mathcal{A}_\phi = \bigoplus_{\mathcal{O}(\mathcal{D})}([g],\rho) \boxtimes ([h],\pi)^{\oplus \delta_{[g],[\phi(h)]}\cdot[\phi^*\rho:\bar{\pi}]}.
\end{equation}
(Here we have suppressed the equivalence \(\hat{H_\phi}\simeq \hat{H_\phi^*}\) that appears during the derivation, and we also note that \(([h],\pi) \mapsto ([\bar{h}],\bar{\pi})\) is tautologically an automorphism of \(\mathcal{D}_G\).)

\begin{example}
    $1+1d$ $\Rep(S_3) \times S_3$ SPT.
\end{example}
In our notation, $S_3={r,s|r^3=s^2=srsr=1}$. There are eight anyon types in $D(S_3)$. Following the convention in \cite{Beigi_2011}, we label them as $A,B,\dots,H$: $A$, $B$, and $C$ correspond to the identity, sign, and two-dimensional representation, respectively; $D$ and $E$ correspond to the conjugacy class $[s]$ associated with the $+$ and $-$ representations of $\mathbb{Z}_2$, respectively; $F$, $G$, and $H$ correspond to the conjugacy class $[r]$ associated with the $\mathbf{1}$, $\omega$, and $\bar{\omega}$ representations of $\mathbb{Z}_3$, respectively.

\paragraph{$\phi(g)=1$} In this case, the left anyon can only be $A$, $B$, or $C$, and $\rho \circ c$ is simply a direct sum of identity representations. Therefore, they must pair with $A$, $D$, and $E$, with multiplicity equal to the dimension of $\rho$. Thus, $\mathcal{A}_\phi=(A_1\oplus B_1 \oplus 2C_1) \boxtimes (A_2 \oplus D_2 \oplus F_2)$. As will be discussed later, this corresponds to the trivial product phase in physical terms.

\paragraph{$\phi(g)=g$} Here, each anyon can only pair with its antiparticle, i.e., $\mathcal{A}_\phi=A_1\boxtimes A_2 \oplus B_1 \boxtimes B_2 \oplus C_1 \boxtimes C_2 \oplus D_1 \boxtimes D_2 \oplus E_1 \boxtimes E_2 \oplus F_1 \boxtimes F_2 \oplus G_1\boxtimes H_2 \oplus H_1 \boxtimes G_2$. As will be seen in the following, this corresponds to the cluster phase in physics.

\paragraph{$\phi(1,r,r^2)=1,\ \phi(s,sr,sr^2)=s$} In this case, $[s]$ maps to itself while $[1]$ and $[r]$ map to $[1]$. The conjugacy classes $[1]$ and $[s]$ behave similarly to the second case, as $\phi$ preserves their conjugacy classes and centralizers. However, $[r]$ and its centralizer $\mathbb{Z}_3$ are all mapped to $1$, so on the right side only $F$ remains in the algebra, pairing with $A$, $B$, and $C$ with multiplicities equal to their dimensions. Therefore, $\mathcal{A}_\phi=A_1\boxtimes A_2 \oplus B_1 \boxtimes B_2 \oplus C_1 \boxtimes C_2 \oplus D_1 \boxtimes D_2 \oplus E_1 \boxtimes E_2 \oplus(A_1\oplus B_1 \oplus 2C_1) \boxtimes F_2$.

\section{Interpretation in Symmetry TFT picture}\label{Interpretation in physics}
It is natural to ask for the physical meaning of the condensable algebras constructed in Sec.~\ref{Characterize SPT by condensable algebras}. The answer is simple: they describe gapped \emph{domain walls} in the \(G\)-quantum double  topological order.

To make this picture concrete, let us briefly review SymTFT and the ``sandwich'' (or unfolding) construction. Consider a \((2+1)\)D topological order \(\mathcal{T}\) realized on a spacetime of the form \(M\times I\), where \(I=[0,1]\) is an interval. This geometry has two spatial boundaries, \(M\times\{0\}\) and \(M\times\{1\}\), which we call the left and right boundaries, respectively. We place on the left boundary (the ``reference boundary'') the categorical symmetry \(\mathcal{C}\) of interest; physically, \(\mathcal{C}\) also serves as the topological data describing a gapped boundary condition of the bulk \(\mathcal{T}\). By the \((2+1)\)D boundary--bulk correspondence, the bulk topological order is then fixed (up to braided equivalence) to be the Drinfeld center of \(\mathcal{C}\),
\[
\mathcal{T} \simeq \Zc(\mathcal{C}).
\]
If we further impose a (possibly different) gapped boundary condition on the right boundary \(M\times\{1\}\), the entire configuration can be viewed as a \((1+1)\)D system living along \(M\), equipped with symmetry \(\mathcal{C}\), obtained by ``closing'' the bulk between two gapped boundaries.

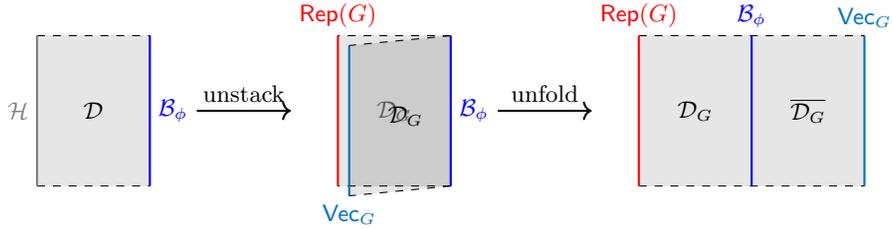
\begin{figure}[h]
    \centering
    \begin{tikzpicture}[scale=0.5]
    % 定义参数
        \def\angle{175} % 折叠角度
        \def\width{3}  % 半平面宽度
        \def\height{4} % 平面高度
        \def\ratio{0.9} % 右半平面缩短比例
        \def\separation{11} % 两个图之间的间距
        
        % 计算旋转后的偏移量 - 175度相当于旋转了几乎180度
        \pgfmathsetmacro{\yoffset}{\width*sin(5)} % 175度 = 180-5度
        
        % 绘制折叠前的原图（左侧）
        \begin{scope}[shift={(-\separation,0)}]
            % 绘制完整平面
            \fill[gray!20] (0,-\height/2) rectangle (\width,\height/2);
            \draw[thick,blue] (\width,-\height/2) -- (\width,\height/2);   % 右边界
            \draw[dashed] (0,-\height/2) -- (\width,-\height/2); % 下边界
            \draw[dashed] (0,\height/2) -- (\width,\height/2);   % 上边界
            \draw[thick, red!50!cyan] (0,-\height/2) -- (0,\height/2);
            
            % 添加标注
            \node[right,blue] at (\width,0) {$\mathcal{B_\phi}$};
            \node[left,red!50!cyan] at (0,0) {$\mathcal{H}$};
            \node at (\width/2,0) {$\mathcal{D}$};
            
        \end{scope}
    
    % 绘制箭头（自动在两个图中间）
        \draw[->, thick] ({-(\separation/2-\width)/2-\separation/2},0) -- ({(\separation/2-\width)/2-\separation/2},0) node[midway,above] {unstack};
    
    % 绘制折叠后的图（右侧）
        \begin{scope}[shift={(0,0)}]
        % 绘制左半平面（保持不动）
            \fill[gray!20] (0,-\height/2) -- (-\width,-\height/2) -- (-\width,\height/2) -- (0,\height/2) -- cycle;
        
        % 绘制右半平面（绕中线旋转175度后的位置，但在x方向上缩短）
            \fill[gray!20] (0,-\height/2) -- 
              (-\width*\ratio,{-\height/2-\yoffset}) -- 
              (-\width*\ratio,{\height/2-\yoffset}) -- 
              (0,\height/2) -- cycle;
        
        % 绘制重叠部分（深灰色）
            \begin{scope}
                \clip (0,-\height/2) -- (-\width,-\height/2) -- (-  \width,\height/2) -- (0,\height/2) -- cycle;
                \fill[gray!40] (0,-\height/2) -- 
                  (-\width*\ratio,{-\height/2-\yoffset}) -- 
                  (-\width*\ratio,{\height/2-\yoffset}) -- 
                  (0,\height/2) -- cycle;
            \end{scope}
        
        % 绘制折痕（中线）
            \draw[thick, blue] (0,-\height/2) -- (0,\height/2);
        
        % 绘制边界线
        % 左半平面边界
            \draw[thick, red] (-\width,-\height/2) -- (-\width,\height/2); % 左边界实线
            \draw[dashed] (0,-\height/2) -- (-\width,-\height/2); % 下边界虚线
            \draw[dashed] (0,\height/2) -- (-\width,\height/2); % 上边界虚线
        
        % 右半平面边界
            \draw[thick, cyan!50!blue] (-\width*\ratio,{-\height/2-         \yoffset}) -- 
                     (-\width*\ratio,{\height/2-\yoffset}); % 右边界实线
            \draw[dashed] (0,-\height/2) -- 
                      (-\width*\ratio,{-\height/2-\yoffset}); % 下边界虚线
            \draw[dashed] (0,\height/2) -- 
                      (-\width*\ratio,{\height/2-\yoffset}); % 上边界虚线
        
        % 添加标注
            \node[right, blue] at (0,0) {$\mathcal{B_\phi}$};
            \node[gray!70!black] at (-1.5,0) {$\mathcal{D}_G$};
            \node[] at (-\width*0.4,{-\yoffset/2}) {$\mathcal{D}_G$};
            \node[below,cyan!50!blue] at (-\width*\ratio,{-\height/2-\yoffset}) {$\Vecc_G$};
            \node[above,red] at (-\width,\height/2) {$\cRep(G)$};
        \end{scope}

        \draw[->, thick] ({(\separation-2*\width)/4},0) -- ({3*(\separation-2*\width)/4},0) node[midway,above] {unfold};

        \begin{scope}[shift={(\separation,0)}]
        % 绘制完整平面
            \fill[gray!20] (-2*\width,-\height/2) rectangle (0,\height/2);
        
        % 绘制边界线
            \draw[thick, cyan!50!blue] (0,-\height/2) -- (0,\height/2);
            \draw[thick,blue] (-\width,-\height/2) -- (-    \width,\height/2); 
            \draw[dashed] (-2*\width,-\height/2) -- (0,-\height/2); 
            \draw[dashed] (-2*\width,\height/2) -- (0,\height/2);   
            \draw[thick, red] (-2*\width,-\height/2) -- (-2*\width,\height/2);
        
        % 添加标注
            \node[above,blue] at (-\width,\height/2) {$\mathcal{B_\phi}$};
            \node at (-3*\width/2,0) {$\mathcal{D}_G$};
            \node at (-\width/2,0) {$\overline{\mathcal{D}_G}$};
            \node[above,red] at (-2*\width,\height/2) {$\cRep(G)$};
            \node[above,cyan!50!blue] at (0,\height/2) {$\Vecc_G$};
        \end{scope}
    \end{tikzpicture}
    \caption{the unfolding trick}
    \label{fig:unfold}
\end{figure}

We now apply this general picture to our setting. In our construction, the reference boundary is \(\mathcal{H}\) and the bulk is
\[
\Zc(\mathcal{H})=\mathcal{D},
\]
and both are \emph{stacked} (i.e.\ layered) topological orders. This allows us to separate the stack into two layers and then perform the standard unfolding move across the \emph{physical} boundary, which in our case is the condensed boundary \(\mathcal{D}_{\mathcal{A}_\phi}\). After unfolding, the bulk degrees of freedom reorganize into a single interface geometry: the left half supports \(\mathcal{D}_G\), while the right half supports \(\overline{\mathcal{D}}_G\) (the time-reversed/chirality-flipped copy). Their outer boundaries are the familiar gapped boundaries \(\Vecc_G\) and \(\cRep(G)\), respectively. In this unfolded picture, the original physical boundary condition \(\mathcal{D}_{\mathcal{A}_\phi}\) is reinterpreted as a \emph{gapped domain wall} between \(\mathcal{D}_G\) and \(\overline{\mathcal{D}}_G\). We denote this domain wall by \(\mathcal{B}_\phi\) (see Fig.~\ref{fig:unfold}).

When an anyon \(([g],\pi)\) crosses the domain wall \(\mathcal{B}_\phi\) from right to left, its flux label (conjugacy class) is mapped as
\[
[g]\longmapsto [\phi(\bar g)].
\]
Conversely, when it crosses from left to right, its charge label transforms by the pullback
\[
\pi \longmapsto \phi^*\bar{\pi},
\]
in agreement with the pairing rule in Eq.~\ref{pair}. The appearance of complex conjugation has a clear physical origin: the right half-space is \(\overline{\mathcal{D}}_G\), i.e.\ the orientation-reversed (time-reversed) copy of \(\mathcal{D}_G\), so charge labels are conjugated when transported across the interface. (Equivalently, one may use the canonical braided equivalence \(\mathcal{D}_G\simeq \overline{\mathcal{D}}_G\) given by \(([g],\pi)\mapsto ([g],\pi^*)\).)

Moreover, if we dimensionally reduce the interface geometry—bringing the two outer boundaries close enough that the sandwiched \((2+1)\)D region can be regarded as ``thin''—the resulting effective theory is a \((1+1)\)D gapped system in the vacuum sector. In this reduced picture, the domain wall data precisely encodes a \((1+1)\)D SPT phase protected by \(\Rep(G)\times G\).

With this physical interpretation in hand, the lattice realization introduced earlier (based on the Kitaev quantum double model) provides an explicit microscopic implementation of the same domain wall. 
To connect this microscopic picture to the categorical computation, consider two ribbon operators meeting at the wall: a creation operator \(W_{q_2k,q_1i}^{([g],\rho)}(\text{path}_1)\) coming from the smooth side and ending on the domain wall, and another creation operator \(W_{pl,p_1j}^{([h],\pi)}(\text{path}_2)\) coming from the rough side and ending at the same site. The question is when there exists a linear combination of their intersecting components,
\[
\sum_{q_2k,\,pl} f_{q_2k,pl}\;
W_{q_2k,q_1i}^{([g],\rho)}(\text{path}_1)\,
W_{pl,p_1j}^{([h],\pi)}(\text{path}_2),
\]
that commutes with all modified local vertex/plaquette operators supported on the wall. Imposing this commutativity condition yields exactly the same constraint as Eq.~\ref{F eq}. In other words, the lattice ``gluing'' condition for ribbon operators at the interface reproduces the bimodule/half-braiding constraint that determines the algebra \(\mathcal{A}_\phi\), thereby matching the lattice domain wall \(\mathcal{B}_\phi\) with the condensable-algebra construction.

\section{discussion}\label{discussion}
Since our construction only produces the \emph{intrinsically mixed} \(G\times \Rep(G)\) SPT phases, it is natural to ask how far these results can be generalized.

\paragraph{Replacing \(\Rep(G)\) by \(\Rep(G')\).}
A first and rather direct extension is to consider symmetry of the form \(G\times \Rep(G')\), where \(G\) and \(G'\) are two (not necessarily isomorphic) finite groups. In this situation the group homomorphism that controls the wall/condensable-algebra data should be replaced by
\[
\phi\in \hom(G,G'),
\]
and the derivations in the previous sections go through with only notational changes: the flux sector is still governed by conjugacy data in \(G\), while the charge sector is governed by representations of \(G'\), with \(\phi\) mediating how the two parts are glued across the interface. We therefore expect an analogous pairing/multiplicity rule, with pullback \(\phi^*\) acting on \(\Rep(G')\).

\paragraph{Toward a full classification of \(G\times \Rep(G)\) SPT phases.}
A more ambitious problem is the complete classification of \emph{all} \(G\times \Rep(G)\) SPT phases, not only the intrinsically mixed subclass. Here we have to be candid: the situation becomes substantially richer, and a uniform closed-form classification seems harder than one might expect.

The key point is that, in Sec.~\ref{Rep(G) times G-module structures on Vecc}, the intrinsically mixed assumption drastically restricts the possible choices of the condensable algebra \(\hat H\). Once this constraint is dropped, there is a much larger family of admissible subalgebras, and correspondingly many more module structures on \(\Vecc\). Even before mixing is addressed, one must understand the independent \(\Rep(G)\) and \(G\) SPT building blocks.

On the \(\Rep(G)\) side, it is known that \((1+1)\)D \(\Rep(G)\)-SPT phases (equivalently, indecomposable \(\Rep(G)\)-module structures on \(\Vecc\)) are characterized by a pair \((L,\psi)\), where \(L\le G\) is a subgroup and \(\psi\in Z^2(L,U(1))\) is a 2-cocycle such that the twisted group algebra \(\mathbb{C}_\psi[L]\) is simple. Equivalently, the corresponding twisted representation category \(\Rep_\psi(L)\) acts trivially in the sense that
\[
\cRep_\psi(L)\simeq \Vecc.
\]
On the \(G\) side, ordinary \((1+1)\)D \(G\)-SPT phases are classified by \(\omega\in Z^2(G,U(1))\), giving the twisted group algebra \(\mathbb{C}_\omega[G]\).

For a \emph{non-mixed} (decoupled) phase obtained by stacking a \(\Rep(G)\)-SPT labeled by \((L,\psi)\) with a \(G\)-SPT labeled by \(\omega\), one may take the condensable algebra to factorize as
\[
\hat H \;\cong\; \mathbb{C}_\psi[L]\;\otimes\;\mathbb{C}_\omega[G],
\]
so the discussion essentially reduces to the two independent classifications.
The genuinely mixed phases are more subtle. Two obstructions appear.

\emph{First}, even the \(\Rep(G)\)-SPT input \((L,\psi)\) is not always easy to enumerate in a uniform way: determining all pairs \((L,\psi)\) for which \(\mathbb{C}_\psi[L]\) is simple can be nontrivial and is often handled case-by-case in practice.

\emph{Second}, once mixing is allowed, the relevant subgroup data underlying \(H\) is no longer a direct product, but rather a ``fiber product'' (or graph-like) subgroup specified by
\begin{equation*}
    N\trianglelefteq G,\qquad L\trianglelefteq A\subseteq G,
\end{equation*}
where \(A\) is a subgroup of \(G\), and \(N\) and \(L\) are normal in \(G\) and \(A\), respectively, such that there is a common quotient group \(P\) with
\begin{equation*}
    G/N \;\simeq\; A/L \;\simeq\; P.
\end{equation*}
In terms of cosets labeled by \(p\in P\), one may write schematically
\begin{equation}
    H \;\equiv\; \bigcup_{p\in P} pN \,\otimes\, pL.
\end{equation}
This form reflects the fact that the \(G\) and \(\Rep(G)\) components are correlated through the shared quotient \(P\), which is precisely what ``mixing'' means at the level of symmetry defects/domain-wall data.

The remaining challenge is then to incorporate twisting data consistently. In the factorized case one may independently insert cocycles \(\psi\) and \(\omega\) into the two tensor factors. For a fiber-product type \(H\), however, there is in general \emph{no canonical} way to embed a given pair of twists \((\psi,\omega)\) into a single associative algebra structure on \(H\): one expects additional compatibility conditions and possibly extra cohomological data controlling how the two twists are coupled along the common quotient \(P\). When these couplings are forced to be trivial (for instance, when \(P\), \(A\), and \(L\) degenerate in the above construction), one recovers the intrinsically mixed family studied in this work.

A systematic classification of these mixed twists—i.e.\ a practical set of invariants and a uniform construction of the corresponding condensable algebras/domain walls—goes beyond the scope of the present paper. We leave a full treatment of the general \(G\times \Rep(G)\) case to future work.

\newpage
\bibliographystyle{unsrt}
\bibliography{ref}
\end{document}